\documentclass[useAMS,usenatbib,onecolumn]{mn2e}
\pdfoutput=1
\usepackage{aas_macros}
\usepackage{epsfig}
\usepackage{amsmath}
\usepackage{amssymb}
\usepackage{leftidx}
\usepackage{natbib}
\usepackage{xcolor}
\usepackage{hyperref}
\hypersetup{colorlinks=true,citecolor=blue,linkcolor=violet}
\def\beq{\begin{equation}}
\def\eeq{\end{equation}}
\def\ba{\begin{eqnarray}}
\def\ea{\end{eqnarray}}
\def\go{\mathrel{\raise.3ex\hbox{$>$}\mkern-14mu
             \lower0.6ex\hbox{$\sim$}}}
\def\lo{\mathrel{\raise.3ex\hbox{$<$}\mkern-14mu
             \lower0.6ex\hbox{$\sim$}}}
\def\bxi{{\mbox{\boldmath $\xi$}}}
\def\bnab{{\mbox{\boldmath $\nabla$}}}

\begin{document}

\title[Heartbeat Stars]{Heartbeat Stars, Tidally Excited Oscillations, and Resonance Locking}
\author[J. Fuller]
{Jim Fuller$^{1,2}$\thanks{Email:
jfuller@caltech.edu}
\\ $^1$ TAPIR, Walter Burke Institute for Theoretical Physics, Mailcode 350-17, California Institute of Technology, Pasadena, CA 91125, USA
\\ $^2$ Kavli Institute for Theoretical Physics, Kohn Hall, University of California, Santa Barbara, CA 93106, USA}

\label{firstpage}
\maketitle

\begin{abstract}

Heartbeat stars are eccentric binary stars in short period orbits whose light curves are shaped by tidal distortion, reflection, and Doppler beaming. Some heartbeat stars exhibit tidally excited oscillations and present new opportunities for understanding the physics of tidal dissipation within stars. We present detailed methods to compute the forced amplitudes, frequencies, and phases of tidally excited oscillations in eccentric binary systems. Our methods i) factor out the equilibrium tide for easier comparison with observations, ii) account for rotation using the traditional approximation, iii) incorporate non-adiabatic effects to reliably compute surface luminosity perturbations, iv) allow for spin-orbit misalignment, and v) correctly sum over contributions from many oscillation modes. We also discuss why tidally excited oscillations are more visible in hot stars with surface temperatures $T \! \gtrsim \! 6500 \, {\rm K}$, and we derive some basic probability theory that can be used to compare models with data in a statistical manner. Application of this theory to heartbeat systems can be used to determine whether observed tidally excited oscillations can be explained by chance resonances with stellar oscillation modes, or whether a resonance locking process is operating.

\end{abstract}


\section{Introduction}
\label{intro}

Heartbeat stars are a growing class of eccentric ($e \gtrsim 0.3$) binary stars in short period ($1 \, {\rm day} \lesssim P \lesssim 1 \, {\rm yr}$) orbits, whose light curves are shaped by tidal distortion, reflection, and Doppler beaming. These effects are most prominent near periastron, combining to generate the characteristic ``heartbeat" signature (whose shape resembles an EKG diagram) that is their namesake. Many heartbeat stars oscillate throughout their orbit due to the tidal excitation of stellar oscillation modes. The smoking gun signature of tidally excited oscillations (TEOs) is that they occur at {\it exact} integer multiples of the orbital frequency. Typical photometric variations of $\Delta L/L \lesssim 10^{-3}$ and characteristic time scales of days have hindered ground-based observations, and only a few heartbeat stars had been discovered in the pre-{\it Kepler} era (see \citealt{decat:00,willems:02,handler:02,maceroni:09}). 

{\it Kepler} has revolutionized studies of heartbeat stars by providing continuous, high precision photometry throughout multiple stellar orbits. The prototypical heartbeat star, KOI-54, has been examined in a series of papers (\citealt{welsh:11,fullerkoi54:12,burkart:12,oleary:14}). It consists of two eccentric A-type stars in a highly eccentric ($e=0.83$), 42 day orbit. KOI-54 exhibits dozens of TEOs, the largest of which occur at exactly 90 and 91 times the orbital frequency. More recently, \cite{thompson:12} presented 17 additional heartbeat stars with a variety of stellar components, and \cite{beck:14} examined 15 sub-giant/red giant heartbeat systems. Additional heartbeat systems include those analyzed in \cite{hambleton:13}, \cite{maceroni:14}, and \cite{hambleton:16}. The radial velocity curves of many heartbeat systems have been measured in \cite{smullen:15,shporer:16,dimitrov:17}, confirming that heartbeat stars are generally near the upper envelope of the binary eccentricity distribution. Currently, over 150 heartbeat stars have been identified in {\it Kepler} data \cite{kirk:16}, although few have been analyzed in detail. 

Despite the fact that many heartbeat stars pulsate, their asteroseismic utility is currently somewhat limited. One reason is that TEOs pulsate at integer harmonics of the orbital frequency (rather than at stellar oscillation mode frequencies), with the largest amplitude TEOs occurring via resonances between orbital harmonics and mode frequencies. Although it may be possible to use this information to perform ``tidal asteroseismology" \citep{burkart:12}, this procedure is difficult. Nonetheless, TEOs differ from self-excited and stochastically excited pulsations because their amplitudes and phases can be straightforwardly predicted from linear theory and compared with observations. Moreover, TEOs allow for the observation of oscillation modes that are normally undetectable, allowing for new opportunities to constrain pulsation physics.

Heartbeat stars also offer an unprecedented opportunity to study tidal interactions between stars, and systems displaying TEOs are especially useful. Given the properties of the host star, the frequency and amplitude of a TEO can be used to identify the stellar eigenmode responsible for the oscillation and the energy contained in the pulsation. With an estimate of the mode damping rate, one can then calculate a tidal dissipation rate and orbital circularization time scale. Many heartbeat stars contain A-F type stellar components, which contain neither a thick convective envelope conducive to damping of the equilibrium tide, nor a large convective core conducive to excitation of the dynamical tide. Hence, A-F stars are not well described by commonly used tidal theories (e.g., \citealt{zahn:77}). Instead, tidal dissipation likely occurs through excitation and damping of TEOs, and so the information contained in heartbeat stars is of great importance.

Previous theoretical work on tidal excitation of stellar oscillations is extensive, and includes \cite{zahn:70,zahn:75,zahn:77,goldreich:89,kumar:95,lai:97,smeyers:98,willems:02,willems:03,willems:03b}, but very few of these works accurately calculate luminosity fluctuations produced by TEOs. One exception is \cite{pfahl:08}, which provides constructive insight on luminosity fluctuations in different types of stars. Unfortunately, the studies above could not compare observed TEOs with theoretical expectations, as TEOs were very difficult to observe in the pre-{\it Kepler} era. Nonetheless, these works contain the foundations of tidal theory upon which our work is built.

In this paper, we make detailed theoretical predictions for photometric amplitudes and phases of TEOs, and the tidal dissipation they produce. In particular, we discuss the relationship between the observed photometric amplitude of a TEO and the energy contained within the stellar pulsation, which is quite sensitive to the stellar structure. Since TEOs are typically due to resonances with gravity modes (g modes), they are usually only detectable in stars without thick convective envelopes (although see \citealt{fullerhd18:13} for an exception). TEOs are also more likely to be observed (although not exclusively found) in high eccentricity systems, where tidal forcing occurs across a wide range of frequencies, and resonances with stellar oscillation modes are more probable. We construct some basic probability theory regarding expectations for the luminosity fluctuations produced by TEOs. Observed systems which defy theoretical expectations can then be used to improve theoretical understanding of tidal interactions in eccentric binaries. In two companion papers \cite{hambleton:17,fullerkic81:17}, we compare these theories with data for KIC 8164262. We measure the parameters of the system and the frequencies and amplitudes of its TEOs, showing that most of them can be explained by chance resonances with g modes, with the exception of its highest amplitude pulsation, which can instead by explained by resonance locking.

Our paper is organized as follows. Section \ref{modes} discusses the basic theory of TEOs, and Section \ref{gmodes} discusses properties of g modes favorable to producing observable TEOs. Section \ref{prob} derives statistical properties of TEOs to be compared with observations. Resonance locking is discussed in Section \ref{reslock}. Tedious calculations required for precise predictions of TEO amplitudes and phases are provided in Sections \ref{dyn}, \ref{sum}, \ref{misalignment}, and \ref{trad}. We conclude and provide discussion in Section \ref{disc}.

\begin{figure*}
\begin{center}
\includegraphics[scale=0.45]{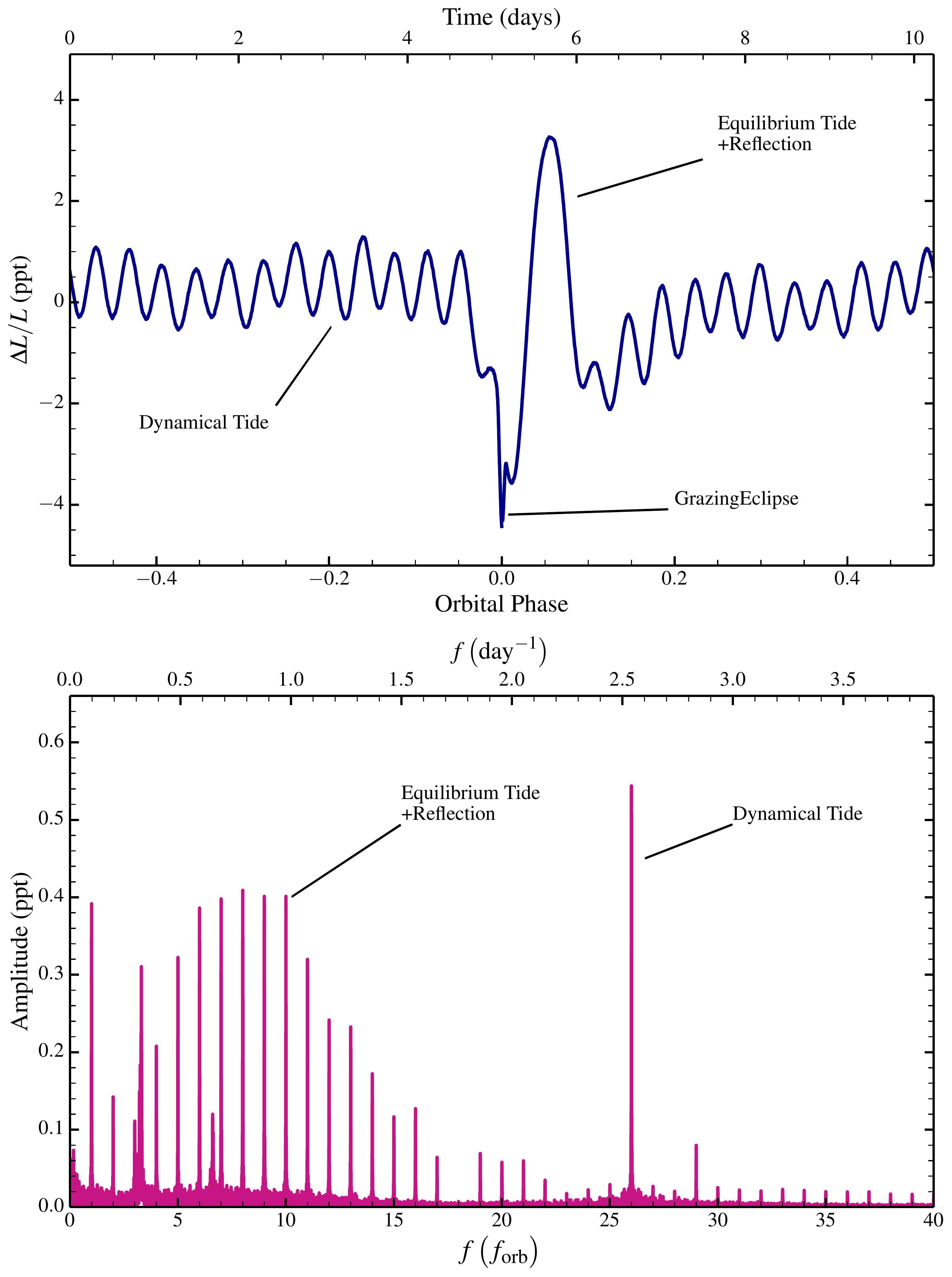}
\caption{ \label{KIC87lc}  {\bf Top:} Phased lightcurve of a heartbeat star with a tidally excited oscillation (KIC 8719324). The sharp variation near periastron (orbital phase=0) is the ``heartbeat" signal produced by the equilibrium tide, reflection, and Doppler boosting. The oscillation away from periastron is produced by a tidally excited oscillation, i.e., the dynamical tide. {\bf Bottom:} Fourier transform of the complete {\it Kepler} lightcurve. The series of evenly spaced peaks are located at integer harmonics of the orbital frequency, and are generated primarily by the heartbeat signal in the light curve. The large amplitude peak at $f \simeq 2.5\,{\rm d}^{-1}$ is produced by the dominant tidally excited oscillation at exactly 26 times the orbital frequency. }
\end{center} 
\end{figure*}

\section{Tidally Excited Oscillations}
\label{modes}

Heartbeat stars are distorted by the time variable tidal potential of the companion star. The response of the star can generally be decomposed into two components: the equilibrium tide and the dynamical tide. The equilibrium tidal distortion is simply the hydrostatic deformation of the star due to the companion, i.e., the steady state distortion that would be produced in the absence of orbital motion. The dynamical tide is the additional non-hydrostatic oscillation of the star that is produced due to the time variable nature of the tidal forcing. The equilibrium tide contributes to the ``heartbeat" signature near periastron, while the dynamical tide is composed of TEOs that are visible at all orbital phases.

Figure \ref{KIC87lc} shows the phased lightcurve of a representative heartbeat star. We point out the periastron heartbeat signature (produced by the equilibrium tide, in addition to reflection and Doppler boosting), and a TEO produced by the dynamical tide. The Fourier transform reveals a characteristic comb of peaks at integer harmonics of the orbital frequency which are produced by the heartbeat distortion in the lightcurve, similar to the comb of peaks produced in the FFT of an eclipse light curve. Some peaks at orbital harmonics are much higher (or lower) amplitude than surrounding peaks, indicating that they are produced primarily by the dynamical tide. Finally, peaks that are not at orbital harmonics may be produced by heat-driven oscillation modes (see, e.g., \citealt{hambleton:13}), non-linear tidal effects (\citealt{fullerkoi54:12,burkart:12,hambleton:13,borkovits:14}), or rotational variation due to spots.

\subsection{Dynamical Tide}

We calculate the dynamical tidal response of the star by decomposing it into the response of each stellar oscillation mode, indexed by $\alpha$. The total stellar response is found by summing over all modes. Each mode contributes to both the equilibrium tide and the dynamical tide, and thus has an equilibrium amplitude $a_{\rm eq}$ and a dynamical amplitude $a_{\rm dyn}$. Binary light curve modeling techniques generally solve for only the equilibrium tidal distortion, because they compute only the hydrostatic distortion produced by the companion. Therefore, after a binary lightcurve model is subtracted away, the amplitude of the remaining oscillations corresponds to dynamical mode amplitudes, $a_{\rm dyn}$. Our calculations below thus focus on the dynamical component of each mode. In Section \ref{dyn}, we formally define the equilibrium mode amplitude, describe how to compute it, and how to calculate the dynamical component of each oscillation mode.

In an eccentric binary, each stellar oscillation mode is forced at every integer harmonic $N$ of the orbital frequency. Consequently, the observable tidal response of a star in an eccentric binary consists of fluctuations at every harmonic of the orbital frequency, with each harmonic composed of a sum over all the star's oscillation modes. In Section \ref{sum}, we describe how to rigorously compute these sums, which must be done carefully to obtain accurate results.

\subsection{Simple Cases}

High amplitude TEOs are usually produced by a near-resonance between an orbital harmonic and a mode frequency. In this case, the luminosity fluctuation at an orbital harmonic $N$ can be approximated by a single term in the sum of equation \ref{AN}. This approach is also described in \cite{fullerkoi54:12,burkart:12}. Here, we provide general formulae including rotation and non-adiabatic effects. When a single resonant mode (indexed by $\alpha$) of azimuthal number $m$ dominates the dynamical tidal response at an orbital harmonic $N$, it produces a sinusoidal luminosity fluctuation of form
\beq
\label{dlumres}
\frac{\Delta L_{N}}{L} \simeq A_N \sin(N \Omega t + \Delta_N)
\eeq
with amplitude
\beq
\label{Aalpha}
A_N = \epsilon_l X_{Nm} V_{lm} \big| Q_{\alpha} L_{\alpha} \big| \frac{\omega_{Nm}}{\sqrt{(\omega_{\alpha} - \omega_{Nm})^2 + \gamma_{\alpha}^2 }} \, .
\eeq
In equation \ref{dlumres}, $\Omega$ is the angular orbital frequency and $t$ is time measured from $t=0$ at periastron. The dimensionless tidal forcing amplitude $\epsilon_l$ is determined by the mass and semi-major axis of the perturber (see equation \ref{epslm}),
\beq
\epsilon_l = \frac{M'}{M} \bigg( \frac{R}{a} \bigg)^{l+1} \, .
\eeq
Here, $M'$ is the mass of the companion, $M$ is mass of the primary, $R$ is its radius,  and $a$ is the semi-major axis of the orbit. The $l$ subscript refers to multipole of the tidal potential which excites the mode, and typically the $l\!=\!2$ components are the most important for tidal excitation. The factor $V_{lm} = |Y_{lm} (i_s,0)|$ (with $Y_{lm}$ a spherical harmonic) accounts for the visibility of the mode given an inclination $i_s$ between the star's rotation axis and the line of sight.

\begin{figure*}
\begin{center}
\includegraphics[scale=0.7]{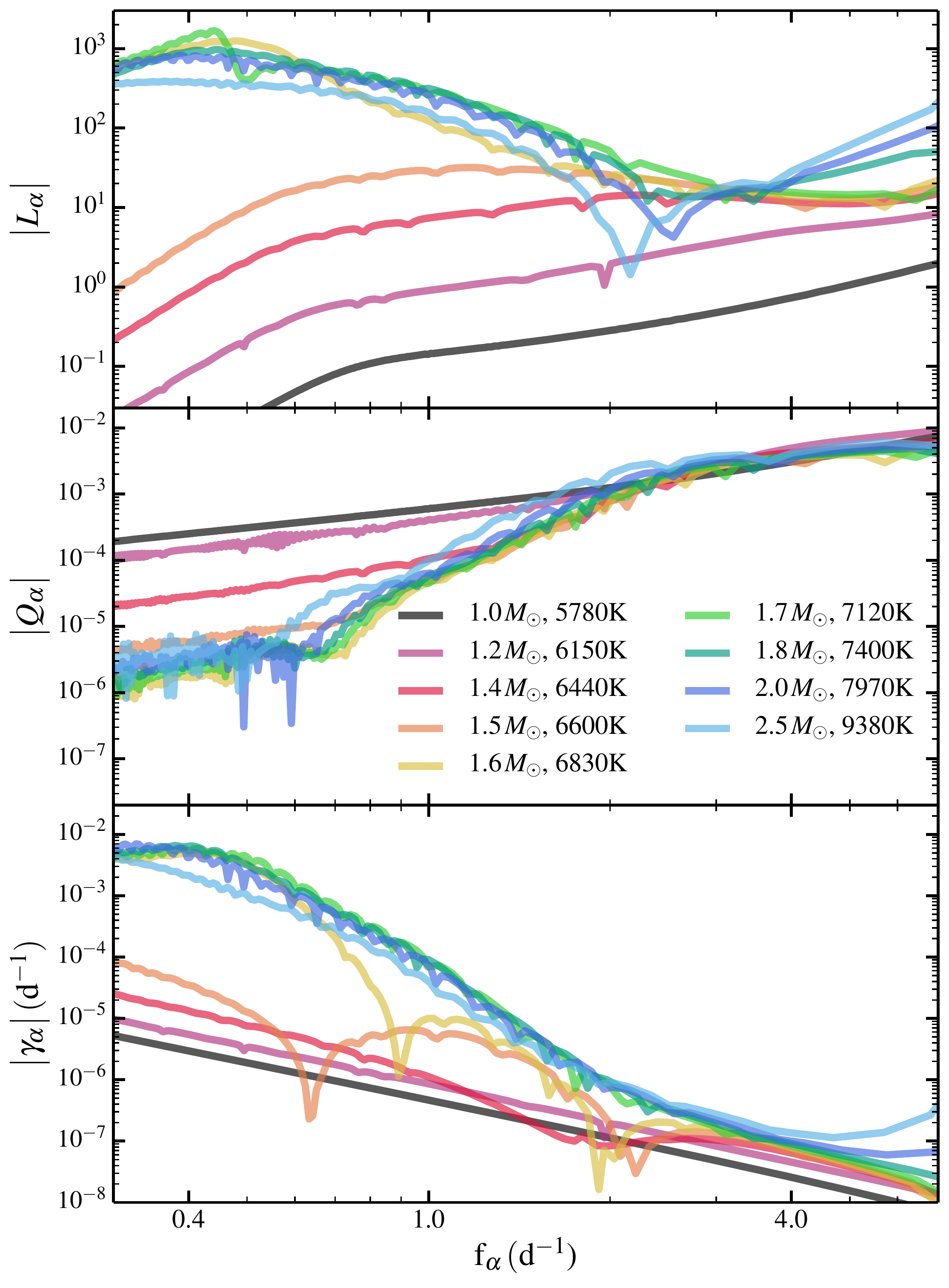}
\end{center} 
\caption{ \label{ModeL} {\bf Top:} Surface luminosity perturbation $L_\alpha$, produced by normalized $l=2$ g modes in $Z=0.02$ main sequence stars of different mass, also labeled by their surface temperature. {\bf Middle:} Mode coupling coefficient $Q_\alpha$ for the same stellar models. {\bf Bottom:} Mode damping rates. Damping times are generally much longer than heartbeat star orbital periods, so that TEO amplitudes do not appreciably decay between successive periastra. The different behavior exhibited by low frequency ($f \lesssim 1 \, {\rm d}^{-1}$) modes in hot stars arises because they propagate close to the photosphere where non-adiabatic effects are large.}
\end{figure*}

$L_\alpha$ describes the observed luminosity fluctuation produced by a mode normalized via equation \ref{modeorthrot}. The value of $L_\alpha$ is sensitive to the stellar model and oscillation mode frequency, and is plotted in Figure \ref{ModeL} for g modes in stars of different masses. $L_\alpha$ is small for g modes in stars with thick convective envelopes ($T_{\rm eff} \lesssim 6500 \, {\rm K}$) because the modes are trapped below the surface by a thick convection zone. In the language of asteroseismology, these modes have very large inertias and are not easily excited. The g modes of hot stars ($T_{\rm eff} \gtrsim 6700 \, {\rm K}$) propagate close to the surface and produce larger luminosity perturbations, hence TEOs are more visible in these stars. The small dips along each curve are created by modes trapped near the convective core, an effect partially washed out by adding small molecular diffusivity to a stellar model. The value of $L_\alpha$ for low frequency modes (${\rm f}_\alpha < 0.8 \, {\rm d}^{-1}$) in stars with $T_{\rm eff} \approx 7000 \, {\rm K}$ is somewhat affected by the treatment of the convective flux perturbation (see Appendix \ref{mesa}), and should be treated cautiously.  The value of $L_\alpha$ is determined primarily by temperature perturbations produced by the mode near the stellar photosphere, thus, a non-adiabatic mode calculation is required for an accurate estimate of $L_\alpha$. 

The dimensionless number $Q_\alpha$ describes the spatial coupling between the stellar oscillation mode $\alpha$ and the tidal potential. It is given by 
\beq
\label{Q}
Q_{\alpha} = \frac{\langle \xi_\alpha | \bnab (r^l Y_{lm}) \rangle}{\omega_\alpha^2}.
\eeq
with all quantities in dimensionless units (i.e., mass in units of $M$, length in units of $R$, time in units of $\sqrt{R^3/GM}$). $Q_\alpha$ can also be expressed in terms of the the surface gravitational potential perturbation associated with each mode (see Section \ref{trad}). Figure \ref{ModeL} shows values of $Q_\alpha$ for g modes in different stars. Typically $Q_\alpha$ is small, of order $10^{-6} \! - \! 10^{-2}$ for gravito-inertial modes using our normalization condition (see Section \ref{dyn}).

\begin{figure*}
\begin{center}
\includegraphics[scale=0.64]{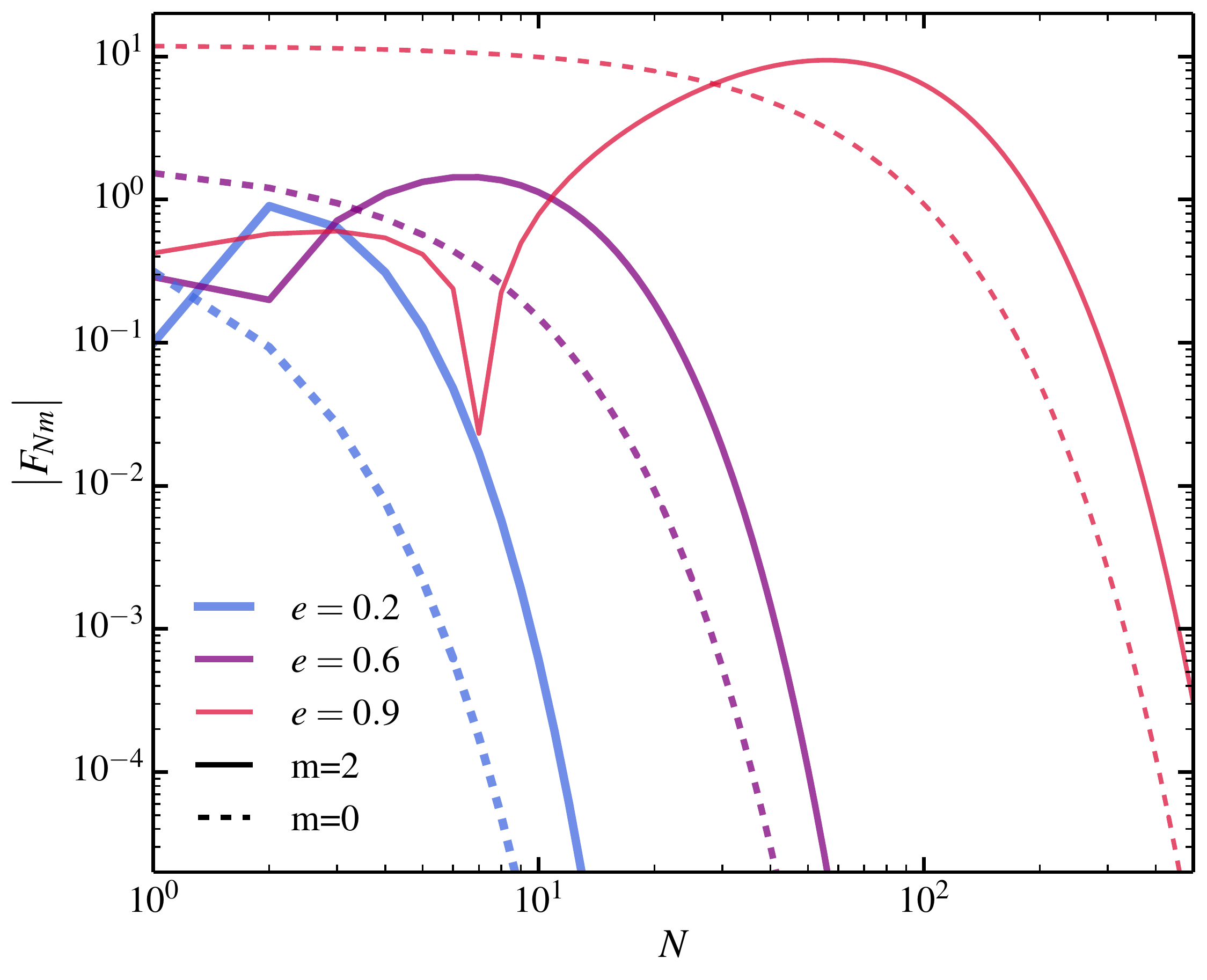}
\end{center} 
\caption{ \label{ModeF} Dimensionless forcing strength $F_{Nm}$ as a function of orbital harmonic $N$ for different orbital eccentricities $e$, plotted for $m=2$ and $m=0$.  }
\end{figure*}

The dimensionless number $X_{Nm}$ describes the strength of the tidal forcing at an orbital harmonic $N$ given an orbit of eccentricity $e$ and is defined in equation \ref{Xnm}.\!\!\footnote{$X_{Nm}$ is similar to a Hansen coefficient, $X^k_{lm}$ (see \citealt{burkart:12}), or a Fourier coefficient, $c_{l,m,k}$, \cite{willems:03}.} Circular and spin-aligned orbits have non-zero values of $X_{Nm}$ only when $N=m$. Misaligned orbits have non-zero values of $X_{Nm}$ for other values of $m$, while higher eccentricity orbits have stronger forcing at higher values of $N$. For aligned spin and orbit, $X_{Nm} = W_{lm} F_{Nm}$, where
\beq
\label{FN}
F_{Nm} = \frac{1}{\pi} \int^\pi_0 d\Psi \frac{\cos \big[N\big(\Psi - e \sin \Psi\big) - m \phi(t)\big]}{\big(1-e \cos \Psi\big)^{l}} \,,
\eeq
and $W_{lm}$ is a constant of order unity defined below equation \ref{epslm}. In equation \ref{FN}, $\Psi$ is the eccentric anomaly, and $\phi(t)$ is the true anomaly. Figure \ref{ModeF} shows $F_{Nm}$ for different orbital eccentricities, for $m=0$ and $m=2$. $F_{Nm}$ peaks at larger orbital harmonics for higher eccentricity, and falls off exponentially at very large $N$, preventing tidal excitation of f modes and p modes. The value of $F_{Nm}$ is very small for $m=-2$, while its value is unimportant for $m=\pm 1$ because $W_{l m}=0$ for $l=2$, $m=\pm 1$.

The last term in equation \ref{Aalpha} is the resonant detuning factor and accounts for the effects of resonant mode excitation. Here, $\omega_\alpha$ is the stellar oscillation mode frequency in the star's rotating frame, $\gamma_\alpha$ is its growth rate, and $\omega_{Nm}$ is the rotating frame tidal forcing frequency,
\beq
\label{omNm}
\omega_{Nm} = N \Omega - m \Omega_s \, ,
\eeq
where $\Omega_s$ is the angular spin frequency. For weakly damped modes, the resonant detuning factor obtains very large values when $\omega_\alpha \simeq \omega_{Nm}$, i.e., near a resonance between the mode frequency and the tidal forcing frequency. We define modes to have time dependence $\propto e^{-i(\omega_\alpha + i \gamma_\alpha) t}$ such that negative values of $\gamma_\alpha$ correspond to damped modes, and positive values of $m$ correspond to prograde modes.

Finally, for aligned spin and orbit, the phase of the luminosity fluctuation relative to periastron is
\beq
\label{Deltaalpha}
\Delta_N = \delta_{\alpha} - m \phi_s \, 
\eeq
where $\delta_\alpha$ is defined in equation \ref{phaselum}, and $\phi_s$ is the azimuthal angle of the observer at periastron. In the misaligned case, the phase is given in equation \ref{dlobs2}. If the mode is nearly adiabatic, then the phase shift is
\beq
\label{deltaalpha}
\delta_{\alpha} \simeq {\rm atan2} \bigg[ \frac{(\omega_{\alpha m} - \omega_{Nm})}{-\gamma_{\alpha m}} \bigg] \, .
\eeq
If the mode is not extremely close to resonance, i.e., $(\omega_\alpha - \omega_{Nm}) \gg \gamma_\alpha$, then $\delta_{\alpha} \simeq \pm \pi/2$.
This is the regime examined in \cite{oleary:14}, in which one may easily compare observed mode amplitudes/phases to theoretical expectations. Note that for a single mode to dominate the tidal response, we must be in the regime where the typical mode spacing, $\Delta \omega_{\alpha m}$, is larger than the typical mode damping rates $\gamma_{\alpha m}$. This criterion is violated for low frequency g-modes with large damping rates, i.e., modes in the traveling wave regime.


\section{Gravity Mode Properties}
\label{gmodes}

Here we examine the g mode properties shown in Figure \ref{ModeL}, and discuss the underlying physics and resulting scaling with mode frequency in different types of stars.

The scaling of $Q_\alpha$ with frequency can be understood as follows. The numerator of equation \ref{Q} is equal to an integral over the density perturabation throughout the star (see equation \ref{Q1}), which is dominated by the integral over the non-oscillatory evanescent tail of the g-mode (see also \citealt{goldreich:99} and \citealt{goodman:98}). For g modes in the WKB limit, the Eulerian density perturbation scales as \citep{luan:17} 
\beq
\label{deltarho}
\delta \rho \sim \frac{\rho r^2}{g H^2} \omega_\alpha^2 \xi_r\, ,
\eeq
where $g$ is gravity and $H$ is the scale height. For WKB g modes, $\xi_r \sim (\omega_\alpha/N) \xi_\perp$, and we have
\beq
\label{deltarho2}
\delta \rho \sim \frac{\rho r^2}{g H^2 N} \omega_\alpha^3 \xi_\perp \, .
\eeq
The length scale of the evanescent tail is $\sim \! H$ and is independent of frequency. Hence, following \cite{luan:17}, the surface gravitational potential perturbation is 
\beq
\label{deltaphi}
\delta \Phi \sim \frac{-4 \pi G}{2 l +1} \frac{\rho r^{l+4}}{g H N R^{l+1}} \omega_\alpha^3 \xi_\perp  \bigg|_{r_{g}} \, .
\eeq
This quantity should be evaluated at the radial location $r_g$ just inside the edge of the g mode propagation cavity, for the evanescent tail that dominates the coupling. For the mode frequencies and stars considered here, this is typically the outer edge of the g mode cavity where $\omega_\alpha=N$ or $\omega_\alpha = L_l$. For low frequency modes in more massive stars, the location of $r_g$ may be just outside the convective core, though we find this is only the case for low frequency modes in the $2.5 \, M_\odot$ star in this study.

\begin{figure*}
\begin{center}
\includegraphics[scale=0.6]{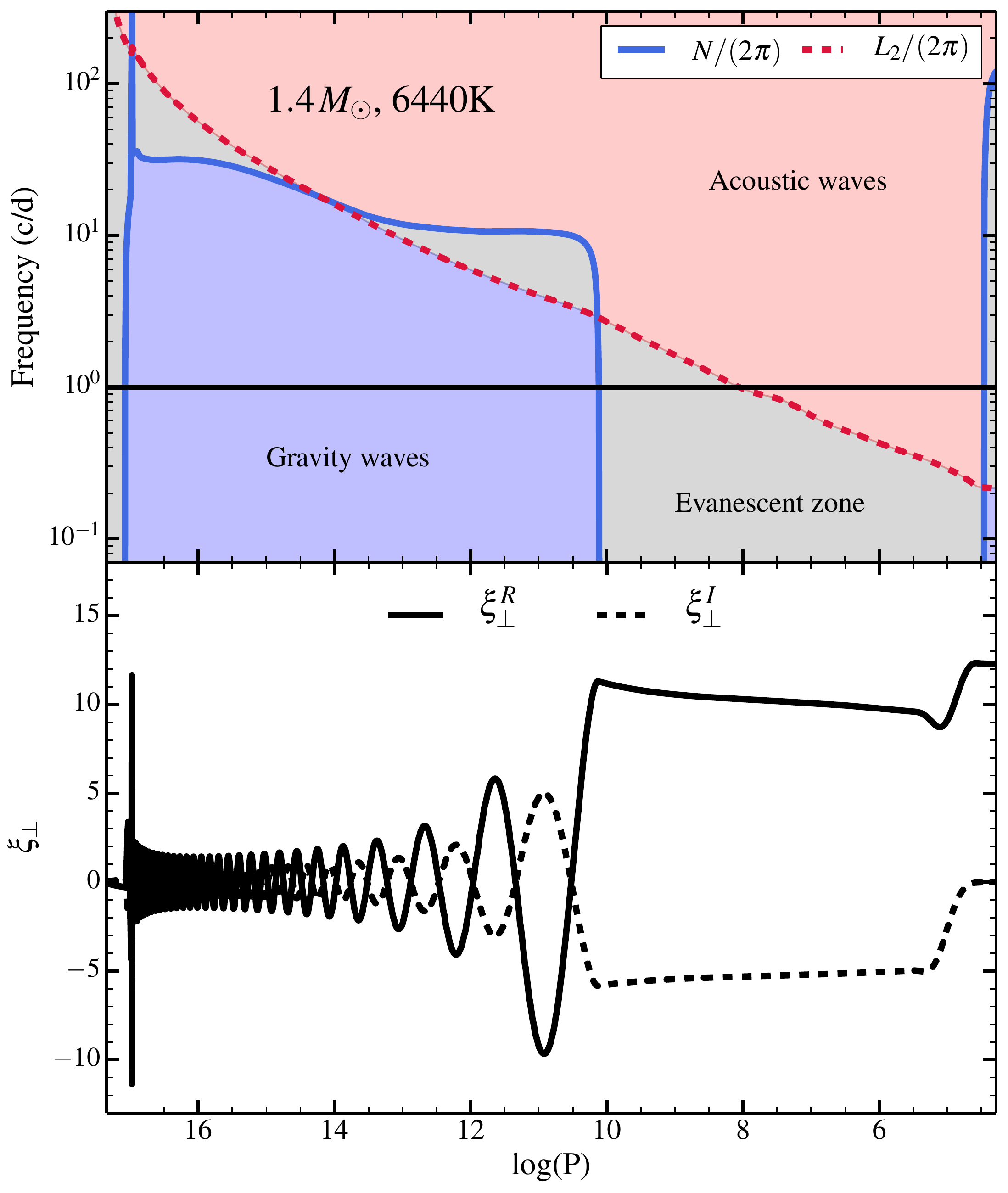}
\caption{ \label{M14}  {\bf Top:} Propagation diagram for a $1.4 \, M_\odot$ ``cool" stellar model, as a function of $\log P$ within the star. For modes with frequencies $f \lesssim 4 \, {\rm c/d}$, the upper edge of the g mode cavity is set by the base of the surface convection zone, and low frequency g modes are separated from the surface by an evanescent zone multiple scale heights in extent. {\bf Bottom:} Real and imaginary components of the normalized horizontal displacement eigenfunction for a g mode with $f_\alpha \simeq 1 \, {\rm c/d}$.}
\end{center} 
\end{figure*}

For our normalization condition, the magnitude of $\xi_\perp$ at a given radius is nearly independent of $\omega_\alpha$. If the mode's turning point $r_g$ is independent of frequency, we find $\delta \Phi \propto \omega_\alpha^3$ and to good approximation $Q_\alpha \propto \omega_\alpha$ (a better scaling taking into account the evanescent wave function is $Q_\alpha \propto \omega^{17/6}$, see \citealt{weinberg:12}, but note their different normalization convention), as can be seen for cool stars in Figure \ref{ModeL}. In cool stars, the location of $r_g$ is the base of the convection zone, nearly independent of mode frequency. Figure \ref{M14} shows a propagation and g mode eigenfunction in a $1.4 \, M_\odot$ where this is this case. In hot stars, the surface convection zone is much thinner, and for modes of $f_\alpha \gtrsim 0.6 \, {\rm d}^{-1}$,  the upper edge of the g mode cavity is determined by the location where $\omega_\alpha = L_l$ rather than the location of the surface convective zone. For this reason, the scaling of $Q_\alpha$ with frequency is steeper, because lower frequency modes have their outer turning point closer to the surface where the density is lower and the mode cannot couple effectively to the tidal potential. 

\begin{figure*}
\begin{center}
\includegraphics[scale=0.6]{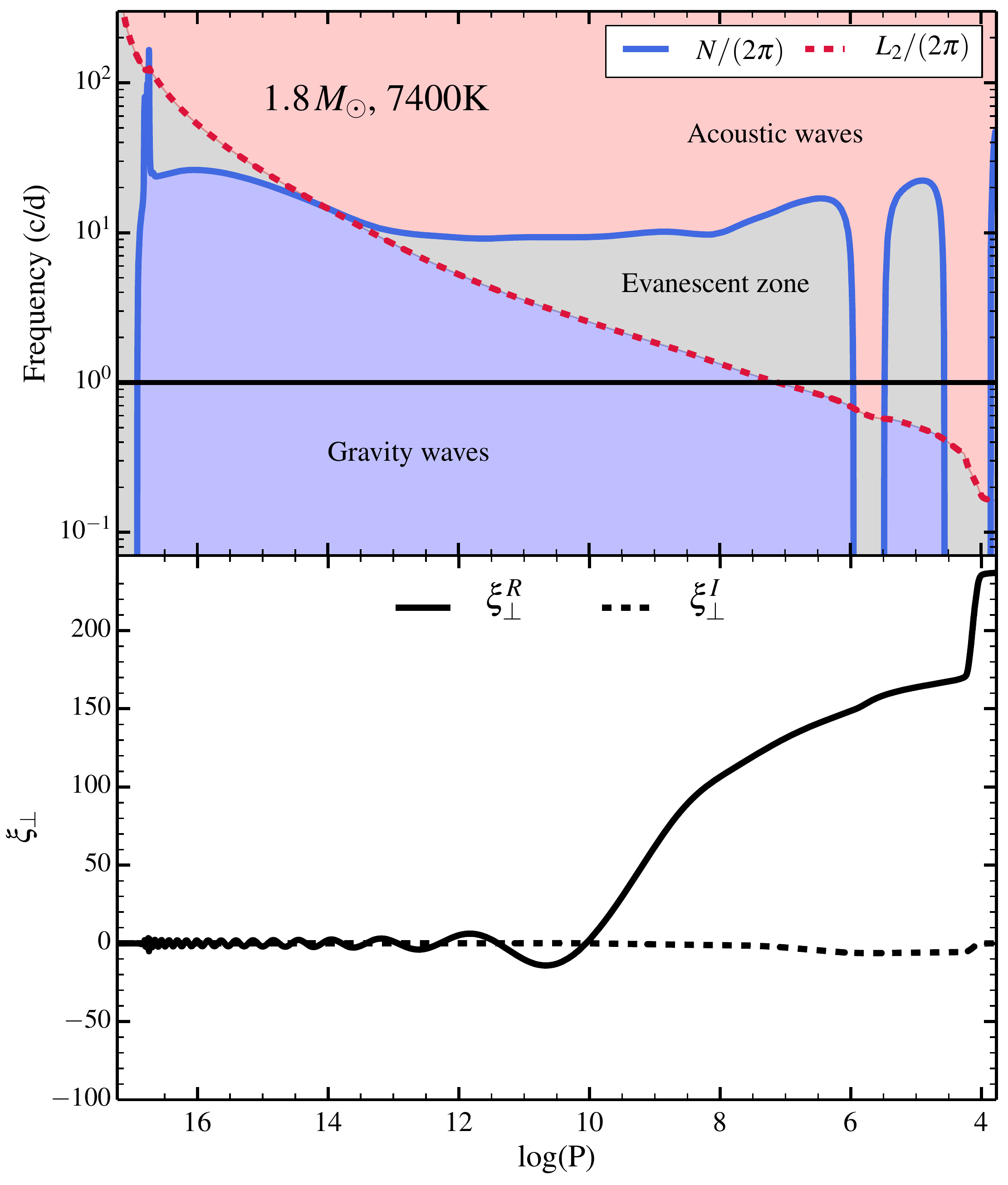}
\caption{ \label{M18} {\bf Top:} Same as Figure \ref{M14}, for a $1.8 \, M_\odot$ ``hot" stellar model. For g modes with frequencies $f \gtrsim 0.6 \, {\rm c/d}$, the upper edge of the g mode cavity is set by the location where $\omega_\alpha = L_l$.  Low frequency g modes are only separated from the surface by a narrow evanescent zone. {\bf Bottom:} Real and imaginary components of the normalized horizontal displacement eigenfunction for a g mode with $f_\alpha \simeq 1 \, {\rm c/d}$. Note the larger $y$-axis scale compared to Figure \ref{M14}. Unlike the mode shown in Figure \ref{M14}, the displacement amplitude grows rapidly for $\log P \lesssim 10$ where the mode propagates into low-density regions.}
\end{center} 
\end{figure*}

The behavior of $L_\alpha$ in Figure \ref{ModeL} can also be understood at a basic level, and is discussed in detail of \cite{pfahl:08}. Following Appendix A2 of \cite{luan:17}, the Lagrangian perturbation to the energy flux for g modes in the WKB limit is of order 
\begin{align}
\label{df}
\frac{ \Delta F}{F} & \sim  \frac{\partial}{\partial r} \xi_r \nonumber \\
& \sim \sqrt{l(l+1)} \frac{\xi_\perp}{r} \, .
\end{align}
with the second line following from the g mode dispersion relation. Equation \ref{df} immediately shows $\Delta F/F \gg \xi_r/r$ for modes in the WKB limit, and temperature/flux perturbations are more important than surface area perturbations in creating the observed luminosity perturbation. To estimate the flux perturbation at the surface of the star, and hence the value of $L_\alpha$, equation \ref{df} should be evaluated at the location where $\omega_\alpha t_{\rm therm} \sim 1$, where $t_{\rm therm}$ is the local thermal time, as discussed in \cite{pfahl:08}. At shallower depths, the flux perturbation is approximately constant. In cool stars, the value of $\Delta F/F$ is small where $\omega_\alpha t_{\rm therm} \sim 1$ because this location is in the convective region, multiple scale heights above the outer boundary of the g mode cavity, where the eigenfunction $\xi_r$ has become much smaller. In hot stars, for g modes with low frequency, this location can be within the g mode cavity, or just above it. Lower frequency modes have their turning point closer to the surface where the density is lower, and hence $\xi_\perp$ is larger. This creates larger values of $L_\alpha$ in hotter stars and for modes with lower frequency. For very low frequency modes ($f_\alpha \lesssim 0.4 \, {\rm d}^{-1}$ in Figure \ref{ModeL}, their amplitude is significantly attenuated by damping because they propagate where $\omega t_{\rm therm} \lesssim 1$, so the value of $L_\alpha$ decreases.

Finally the behavior of $\gamma_\alpha$ is relatively simple to understand. The mode damping rate is
\beq
\label{gamma}
\gamma_\alpha \sim \frac{ \int k_r^2 K |\bxi|^2 dr}{\int |\bxi|^2 dr} \, ,
\eeq
where $K$ is the thermal diffusivity and $k_r \approx \sqrt{l(l+1)} N/(\omega_\alpha r)$ is the g mode radial wavenumber. For g modes with a fixed propagation cavity, we obtain $\gamma_\alpha \propto \omega_\alpha^{-2}$ as seen for g modes in cool stars in Figure \ref{ModeL}. In hot stars, lower frequency modes propagate closer to the surface where $K$ is larger. This creates larger mode damping rates and a steeper scaling with mode frequency. As mentioned above, very low frequency modes ($f_\alpha \lesssim 0.4 \, {\rm d}^{-1}$) propagate into regions where they are strongly damped. In this case, they are essentially damped on a wave crossing time scale, such that $\gamma_ \alpha \sim \int v_g dr \sim \Delta \omega_\alpha \propto \omega_\alpha^2$, and so the mode damping rate starts to decrease with decreasing frequency. Our GYRE calculations only consider radiative diffusion as a source of damping, and neglect interaction with convection, or damping due to convective viscosity. This is problematic for stars near the $\gamma$-Doradus instability strip whose instability is driven by interaction with convection \citep{guzik:00}. Convective turbulent viscosity could potentially be important, but is likely sub-dominant to convective flux blocking effects as discussed in \citep{guzik:00}. 


We comment on the real vs. imaginary pieces of mode eigenfunctions. First, imaginary components of the mode eigenfunctions can be quite large (see Figure \ref{M14}), comparable to the real components, even in somewhat cool stars where non-adiabatic effects are weaker than in hot stars. This generally causes $L_\alpha$ to have a significant imaginary component, and mode luminosity perturbations discussed in Section \ref{lum} can be significantly shifted in phase relative to the expectation for an adiabatic mode. Figures \ref{M14} and \ref{M18} show that $\xi_\perp^I =0$ at the star's surface, but this is merely a consequence of the arbitrary choice of a phase convention $\xi_r^I = 0$ at the surface. Since the outer boundary condition is $\delta P = \rho g \xi_r$ and $\xi_\perp \propto \delta P$, the boundary condition results in $\xi_\perp^I = 0$ a the surface. This is not true for temperature and flux perturbations, which can have a substantial phase shift relative to the mode displacement.

\section{Statistical Approach}
\label{prob}

In practice, accurately predicting amplitudes of TEOs is extremely difficult. The reason is that the mode amplitudes depend sensitively on the detuning factor, $\omega_{Nm}/\sqrt{(\omega_\alpha - \omega_{Nm})^2 + \gamma_\alpha^2} \gg 1$ for nearly resonant modes. Very small changes in the stellar model that alter the mode frequencies $\omega_\alpha$ translate into very large changes in the detuning factor and predicted TEO amplitudes.

A useful approach is thus to examine tidal mode excitation from a statistical point of view. Given a stellar and orbital model, we may reliably calculate the values of the terms in equation \ref{Aalpha} for modes with $\omega_\alpha \approx\omega_{Nm}$. We may then estimate the expected TEO amplitudes from the expected values of the mode detuning factor. Consider tidal forcing frequencies for modes of azimuthal number $m$. The forcing frequencies are uniformly distributed at intervals of $\Delta \omega_{Nm} = N\Omega$ in frequency. The stellar eigenfrequencies, in turn, are distributed with a typical spacing of $\Delta \omega_\alpha$ determined by the stellar structure. Typically these mode frequencies are distributed smoothly so that we may calculate a typical mode frequency separation $\Delta \omega_{\alpha}$ at frequencies where $\omega_\alpha \approx \omega_{Nm}$. 
In heartbeat star systems, we find that both regimes $\Delta \omega_{Nm} < \Delta \omega_{\alpha}$ and $\Delta \omega_{Nm} > \Delta \omega_{\alpha}$ may be realized.

For a forcing frequency $\omega_{Nm}$, we are interested only in the most resonant mode, i.e., the mode for which 
\beq
\label{deltaomN}
\delta \omega_N = |\omega_{Nm} - \omega_\alpha|. 
\eeq
is minimized. The minimum possible frequency difference is 
\beq
\label{dommin}
\delta \omega_{N, {\rm min}} = 0
\eeq
while the maximum possible frequency difference between a forcing frequency and the mode closest to resonance is
\beq
\label{dommax}
\delta \omega_{N, {\rm max}} = \Delta \omega_\alpha/2
\eeq
This is true regardless of whether $\Delta \omega_{Nm} < \Delta \omega_{\alpha}$ or whether $\Delta \omega_{Nm} > \Delta \omega_{\alpha}$. If there is no correlation between mode frequencies and forcing frequencies, the mean frequency difference is 
\beq
\label{dommean}
\delta \omega_{N,{\rm mean}} = \Delta \omega_\alpha/4.
\eeq

For ease in presentation, we define the non-resonant luminosity amplitude
\beq
\label{fancyL}
\mathcal{L}_{N} = \epsilon_{l} X_{Nm} V_{lm} |Q_{\alpha}  L_{\alpha}| \, ,
\eeq
Note that some factors in this equation (e.g., $X_{Nm}$) are defined at forcing frequencies $\omega_{Nm}$, while others (e.g., $Q_\alpha$) are defined at mode frequencies $\omega_\alpha$. In practice, the values of $Q_\alpha$, $\Delta \omega_\alpha$, etc., are sufficiently smooth that one can estimate their values at a forcing frequency by interpolating between neighboring values of $\omega_\alpha$. This process may be more difficult in the sub-inertial regime ($\omega_\alpha < 2 \Omega_s$) where inertial modes make the mode spectrum very dense. 

From equation \ref{dommean}, the most probable value of the luminosity fluctuation produced by a nearly-resonant mode is
\beq
\label{dlummean}
\bigg| \frac{\Delta L_N}{L} \bigg|_{\rm med} \simeq \bigg| 4 \mathcal{L}_{N} \frac{\omega_{Nm}}{\Delta \omega_\alpha} \bigg|.
\eeq
Equation \ref{dlummean} represents the most likely luminosity fluctuation amplitude, assuming the TEO is produced by a single near-resonant oscillation mode. The expected value of the amplitude is somewhat larger than equation \ref{dlummean}, because of the high possible amplitudes attainable very close to resonance. The maximum possible luminosity fluctuation amplitude is
\beq
\label{dlummax}
\bigg| \frac{\Delta L_N}{L} \bigg|_{\rm max} \simeq \bigg| \mathcal{L}_{N} \frac{\omega_{Nm}}{\gamma_\alpha} \bigg|.
\eeq

We can also calculate the probability of a luminosity fluctuation $\Delta L_N/L$ at an orbital harmonic $N$ exceeding some critical value $\Delta L_c/L$, given by
\begin{align}
\label{lc}
{\rm P}\bigg[\frac{\Delta L_N}{L} > \frac{\Delta L_c}{L} \bigg] &= {\rm P} \bigg[  A_N > \frac{\Delta L_c}{L} \bigg]
\end{align}
with $A_N$ given by equation \ref{Aalpha}. A little rearranging yields
\beq
\label{plc}
{\rm P}\bigg[\frac{\Delta L_N}{L} > \frac{\Delta L_c}{L} \bigg] = {\rm P} \bigg[ \delta \omega_N < \sqrt{Z_\alpha} \frac{\Delta \omega_\alpha}{2} \bigg]
\eeq
with 
\beq
\label{zalpha}
Z_\alpha =  \bigg( \frac{\Delta L_c}{L} \bigg)^{-2} \bigg( 2 \mathcal{L}_{N} \frac{\omega_{Nm}}{\Delta \omega_\alpha} \bigg)^2 - \bigg( \frac{2 \gamma_\alpha}{\Delta \omega_\alpha} \bigg)^2.
\eeq
Assuming no correlation between $\omega_{Nm}$ and $\omega_\alpha$, the probability that the frequency detuning $\delta \omega_N$ be less than some fraction $x$ of the maximum detuning $\Delta \omega_\alpha/2$ is simply equal to $x$, as long as $0<x<1$. Then we have
\begin{align}
\label{plc2}
{\rm P}\bigg[\frac{\Delta L_N}{L} > \frac{\Delta L_c}{L} \bigg] &= \sqrt{Z_\alpha} \ \ {\rm for} \ 0 < Z_\alpha < 1 \nonumber \\
&= 0 \ \ {\rm for} \ Z_\alpha < 0 \nonumber \\
&= 1 \ \ {\rm for} \ Z_\alpha > 1.
\end{align}
Then, the expected number of luminosity fluctuations above some amplitude $\Delta L_c/L$, distributed across all orbital harmonics $N$, is
\beq
\label{numex}
{\rm Num}_{\rm ex} \bigg[\frac{\Delta L_N}{L} > \frac{\Delta L_c}{L} \bigg] = \sum_{N_{\rm min}}^{N_{\rm max}} {\rm P}\bigg[\frac{\Delta L_N}{L} > \frac{\Delta L_c}{L} \bigg].
\eeq
In practical calculations, the sum of equation \ref{numex} must be truncated, and the value of $N_{\rm max}$ must be large enough to include all harmonics that can substantially contribute to the sum.
In most cases, very low and high frequency TEOs do not substantially contribute to the sum, so the truncation is not problematic.

It is also useful to calculate the expected density of TEOs as a function of $A_N$. Inverting equation \ref{dlumres}, we have
\beq
\label{domn}
\delta \omega_N = \sqrt{ \omega_{Nm}^2 \big( \mathcal{L}_N/A_N \big)^2- \gamma_\alpha^2 } \, .
\eeq
Differentiation with respect to $A_N$ yields
\beq
\label{ddomn}
d \delta \omega_N = \frac{\omega_{Nm} \big( \mathcal{L}_N/A_N \big)^2 }{\sqrt{\big( \mathcal{L}_N/A_N \big)^2-  \big(\gamma_\alpha/\omega_{Nm} \big)^2 }}  \frac{d A_N}{A_N} \, .
\eeq
Now, for a given orbital harmonic N with a resonant detuning $\delta \omega_N$, the number of TEOs per unit detuning frequency is 
\beq
\label{dndomn}
\frac{d n}{d \delta \omega_N} = \frac{2}{\Delta \omega_\alpha} \, .
\eeq
This can be understood from the fact that around each forcing frequency the most resonant mode always has $0 < \delta \omega_N < \Delta \omega_\alpha/2$. Then we find that the probability density of modes as a function of amplitude is
\beq
\label{dndln}
\frac{d n}{d A_N} = \frac{2 \omega_{Nm}}{\Delta \omega_\alpha} \frac{ \big (\mathcal{L}_N/A_N \big)^2}{\sqrt{\big( \mathcal{L}_N/A_N \big)^2-  \big(\gamma_\alpha/\omega_{Nm} \big)^2 }} \frac{1}{A_N} \, . 
\eeq
With a stellar/orbital model, this expression can be evaluated at each orbital harmonic N, providing a probability density of TEOs as a function of frequency and amplitude. One can then compare the theoretical TEO distribution with the observed distribution.

\section{Tidal Dissipation and Resonance Locking}
\label{reslock}

The angular momentum contained by a mode $\alpha$, together with its complex conjugate (c.f. \citealt{burkart:14}), using our normalization condition in Section \ref{dyn} is
\beq
\label{jmode}
J_\alpha = 2 m \omega_\alpha |a_\alpha|^2 MR^2 \, .
\eeq
Differentiating with respect to time yields $\dot{J}_\alpha = 2 \gamma_\alpha J_\alpha$, and using $\dot{E}_\alpha = (N \Omega/m) \dot{J}_\alpha$, the orbital energy dissipation rate due to a tidally forced mode (and its complex conjugate) is 
\beq
\label{eorbdot}
\dot{E}_{\rm orb,tide} = 4 \omega_\alpha N \Omega \gamma_\alpha |a_\alpha|^2 M R^2.
\eeq
Recalling that $\gamma_\alpha \! < \! 0$ corresponds to a damped mode, $\dot{E}_{\rm orb,tide}$ may be either positive or negative, depending on the sign of $\omega_\alpha$, the mode frequency in the star's rotating reference frame. Physically, this corresponds to the fact that the orbit may lose energy due to a resonance with a prograde mode (positive $\omega_\alpha$) but it may gain energy due to a resonance with a mode that is retrograde in the rotating frame (negative $\omega_\alpha$) but prograde in the inertial frame (positive $N \Omega$).

Using the relation $J_{\rm orb} = -2 \sqrt{1-e^2} E_{\rm orb}/\Omega$, we can solve for the eccentricity evolution of the orbit. The resulting circularization time scale is 
\beq
\label{te}
t_{e} = \frac{-e}{\dot{e}} = \frac{2 e^2}{1-e^2} \frac{N \sqrt{1-e^2}}{N \sqrt{1-e^2} - m} t_{\rm orb} \, ,
\eeq
where the orbital decay timescale is 
\beq
\label{torb}
t_{\rm orb} = \frac{E_{\rm orb}}{\dot{E}_{\rm orb}} \, .
\eeq
For a decaying orbit, $t_{\rm orb}$ is positive. In this case, the eccentricity is damped when $N\sqrt{1-e^2}>m$, but it is excited when $N\sqrt{1-e^2}<m$.

The luminosity fluctuation amplitude produced by a mode (together with its complex conjugate) is
\beq
\label{reslum}
A_N = \big| 2 a_{\alpha} V_{lm} L_\alpha \big| \, .
\eeq The energy dissipation rate of a mode can thus be measured from its observed amplitude,
\beq
\label{eorbdotvis}
\dot{E}_{\rm orb,tide} = \omega_\alpha N \Omega \gamma_\alpha M R^2 \frac{A_N^2}{V_{lm}^2 L_\alpha^2}.
\eeq
Hence, with a good stellar/orbital model such that one may calculate values of $\gamma_\alpha$, $L_\alpha$, etc., inserting observed TEO amplitudes into equation \ref{eorbdotvis} allows one to calculate the tidal energy dissipation rate.

\subsection{Resonance Locking}

TEOs with higher amplitudes than expected from a statistical analysis (Section \ref{prob}) are created by tidally forced modes unusually close to resonance. Such TEOs are good candidates for resonantly locked modes, the process of which is analyzed in detail by \cite{witte:99,fullerkoi54:12,burkart:14}. Here we discuss observational signatures of resonantly locked modes.

A mode is near resonance when one of its eigenfrequencies is approximately equal to a forcing frequency, $\omega_{\alpha} \simeq \omega_{Nm}$,
or in the inertial frame,
\beq
\label{ressig}
\sigma_{\alpha} = \omega_\alpha + m \Omega_s \simeq \sigma_{N} = N \Omega.
\eeq
If the mode is locked in resonance, then this condition does not change with time, i.e.,
\beq
\label{rest}
\dot{\sigma}_{\alpha} \simeq \dot{\sigma}_N = N \dot{\Omega},
\eeq
where the dot denotes the time derivative. 

Mode eigenfrequencies change with time for various reasons (e.g., stellar evolution, tidal spin-up or spin-down, magnetic braking, etc.). Similarly the orbital frequency may change due to several effects (e.g., orbital decay via gravitational waves, tidally induced orbital decay). For heartbeat stars beginning to evolve off the main sequence, we expect stellar evolution and tidal dissipation to be the dominant effects. Assuming the mode frequencies evolve only via stellar evolution and tidal spin-up, while the orbital frequency evolves due only to tidal dissipation, resonance locking requires
\beq
\label{res1}
\frac{\dot{\sigma}_{\alpha,*}}{\sigma_{\alpha}} + \frac{\dot{\sigma}_{\alpha,{\rm tide}}}{\sigma_{\alpha}} = \frac{\dot{\Omega}_{\rm tide}}{\Omega},
\eeq
where the first term on the left hand side accounts for the change in mode frequency due to stellar evolution and the second accounts for frequency change due to tidal torques on the star. For simplicity, we assume the star maintains rigid rotation at all times. The tidal spin term is (see \citealt{burkart:14})
\begin{align}
\dot{\sigma}_{\alpha,{\rm tide}} &= \frac{\partial \sigma_\alpha}{\partial \Omega_s} \dot{\Omega}_{s,{\rm tide}} \nonumber \\
&= m B_{\alpha} \frac{\dot{J}_{\rm tide}}{I} \nonumber\\
&= \frac{m^2 B_{\alpha}}{I N \Omega} \dot{E}_{\rm tide}.
\end{align}
Here, $B_\alpha = (1/m) \partial \sigma_\alpha/\partial \Omega_s$, which evaluates to $B_\alpha \simeq 1 - 1/l(l+1)$ for asymptotic g-modes in the the slowly rotating limit, while $I$ is the star's moment of inertia. 

The energy dissipation rate $\dot{E}_{\rm tide}$ is the tidal energy dissipation rate in the star in the inertial frame and obeys the relation $\dot{E}_{\rm tide} = N \Omega \dot{J}_{\rm tide}$. Energy conservation requires $\dot{E}_{\rm tide} = - \dot{E}_{\rm orb,tide}$, where $E_{\rm orb}=-GMM'/(2a)$ is the orbital energy. The orbital frequency and energy are related by
\beq
\frac{\dot{\Omega}_{\rm tide}}{\Omega} = \frac{3}{2} \frac{\dot{E}_{\rm orb,tide}}{E_{\rm orb}},
\eeq
so the resonance locking condition (equation \ref{rest}) requires
\beq
\label{sigstar}
\frac{\dot{\sigma}_{\alpha,*}}{\sigma_\alpha} = \bigg[\frac{3}{2} + \frac{m^2 B_\alpha E_{\rm orb}}{N^2 I \Omega^2} \bigg] \frac{\dot{E}_{\rm orb,tide}}{E_{\rm orb}}
\eeq
Equation \ref{sigstar} demonstrates a key result: if resonance locking occurs, the orbital energy loss is set by a time scale proportional to $\sigma_\alpha/\dot{\sigma}_{\alpha,*}$, i.e., a stellar evolution time scale. Therefore, if resonance locking occurs, the orbital decay/circularization timescales are set by the rate of stellar evolution. The term in brackets can be positive or negative, and resonantly locked modes could have either increasing or decreasing frequencies.

Inserting equation \ref{eorbdot} into equation \ref{sigstar}, we find the resonantly locked mode amplitude is
\beq
\label{resamp}
|a_\alpha|_{\rm ResLock} = \frac{1}{2} \bigg[ \frac{-N \Omega}{\chi_\alpha \omega_\alpha \gamma_\alpha t_{\alpha, \rm ev}} \bigg]^{1/2}\, ,
\eeq
where we have defined the mode evolution time scale due to stellar evolution,
\beq
\label{sigev}
t_{\alpha,{\rm ev}} = \frac{\sigma_{\alpha}}{\dot{\sigma}_{\alpha,*}} \, 
\eeq
which may be either positive or negative. In equation \ref{resamp}, 
\beq
\label{chi}
\chi_\alpha = \frac{3 N^2 (M+M') R^2}{M' a^2} - \frac{m^2 B_\alpha}{\kappa},
\eeq
with $\kappa$ the dimensionless moment of inertia such that $I = \kappa M R^2$. The value of $\chi_\alpha$ is closely related to the critical harmonic $N_c$ discussed in \cite{fullerkoi54:12} and the moment of inertia ratio $r$ from \cite{burkart:14}. Note that the mode amplitude is only real if the quantity in brackets in equation \ref{resamp} is positive, this is equivalent to the existence of a fixed point as discussed in \cite{burkart:14}.

Equations \ref{reslum} and \ref{resamp} illustrate the purpose of this endeavor: if we know the properties of a system (stellar and orbital parameters, oscillation mode properties, etc.), then we can predict the luminosity fluctuation produced by a resonantly locked mode. Including the resonantly locked mode and its complex conjugate, the explected luminosity variation is
\beq
\label{reslocklum}
A_{\rm ResLock} = \bigg[ \frac{c_\alpha}{\gamma_\alpha t_{\alpha, {\rm ev}}} \bigg]^{1/2} V_\alpha |L_\alpha| \, ,
\eeq
where $c_\alpha = N \Omega/(\omega_\alpha \chi_\alpha)$. Equation \ref{resamp} also provides contstraints on the possible frequencies of resonantly locked modes, as only modes in limited frequency regimes will produce real values of $|a_\alpha|_{\rm lock}$. 

The orbital evolution is easily calculated during a resonance lock. Equation \ref{sigstar} can be rearranged to yield
\beq
\label{eorbdot2}
\dot{E}_{\rm orb,tide} \simeq - \frac{(N \Omega)^2 M R^2}{\chi_\alpha t_{\alpha,{\rm ev}}} \, .
\eeq
When the first term of equation \ref{chi} dominates, this simplifies to 
\beq
\label{eorbdot3}
t_{\rm orb} \simeq \frac{3}{2} t_{\alpha,{\rm ev}} \, ,
\eeq
which is also derived in \cite{fullersattide:16}. Equation \ref{eorbdot3} clearly exhibits that while resonance locking occurs, orbital decay proceeds on the timescale $t_{\alpha,{\rm ev}}$ at which the resonantly locked mode's intrinsic frequency changes. Typically, this is comparable to the stellar evolution timescale. 

Equation \ref{resamp} can be easily generalized to include additional forms of stellar and orbital evolution. Incorporating these terms merely requires the replacement of $t_{\alpha,{\rm ev}}$ with a general evolution timescale $t_{\rm ev}$ given by 
\beq
\label{sigevgen}
t_{\rm ev} = \bigg[ \frac{\dot{\sigma}_{\alpha,*}}{\sigma_{\alpha}} - \frac{\dot{\Omega}_{\rm other}}{\Omega} + \frac{m B_{\alpha} \Omega_s}{\sigma_\alpha} \frac{\dot{\Omega}_{s,\rm other}}{\Omega_s} \bigg]^{-1},
\eeq
where the second and third terms on the right hand side account for additional orbital and spin frequency changes. 

We comment that resonance locks are not always stable, and \cite{burkart:14} presents a detailed analysis of their stability. A key issue is that the above analysis assumes an ``adiabatic approximation" for the mode amplitude, such that the mode amplitude is given by equation \ref{modedyn}. This approximation neglects changes in mode energy, and results in the approximation $\dot{E}_{\rm tide} = -\dot{E}_{\rm orb}$ used above. This approximation is always valid when $\omega_\alpha/(\gamma_\alpha^2 t_{\alpha,{\rm ev}}) < 1$. For typical mode frequencies $f_{\alpha}= 1 \, {\rm c/d}$ and evolution rates $t_{\alpha,{\rm ev}} = 1 \, {\rm Gyr} $, this criterion is satisfied for $\gamma_\alpha \! \gtrsim \! 4 \times 10^{-6} \, {\rm c/d}$. Comparison with Figure \ref{ModeL} shows that the adiabatic approximation is valid for sufficiently low frequency modes in relatively hot stars. It is less likely to be valid for higher frequency g modes and in cool stars with smaller damping rates. When the adiabatic approximation is not valid, resonance locking may still occur, but in a restricted parameter space and with more complex dynamics, as discussed in \cite{burkart:14}. In our companion paper \citep{fullerkic81:17}, we show that modes in KIC8164262 easily satisfy the ``no backreaction" approximation  (equation 53 of \citealt{burkart:14}) even though those modes do not satisfy the strict adiabatic criterion above. The no backreaction approximation allows us to use the adiabatic approximation at the mode amplitude required to sustain a resonance lock, even though the adiabatic approximation could fail at higher amplitudes.

\section{Dynamical Mode Amplitudes}
\label{dyn}

The tidal response of the star can be generally separated into two components: a hydrostatic equilibrium tide and a non-hydrostatic dynamical tide. The equilibrium tide corresponds to the perturbation induced in the star by a static external gravitational potential, and causes the star to distort into an elliptical shape so that hydrostatic equilibrium is maintained. The equilibrium tide is usually responsible for tidal ellipsoidal variations in close binaries, and creates part of the near-periastron luminosity variation observed in heartbeat stars. 

In realistic systems, the external gravitational potential is not static (unless the orbit is both circular and synchronized), and the instantaneous amplitude of the equilibrium tide fluctuates. However, because of the finite inertia of the fluid, the fluctuating equilibrium tide induces a dynamical tidal response which can be extremely important for dissipative tidal processes. To understand the relation between the equilibrium tidal response and the dynamical tidal response, consider the linearized momentum equation in the star's rotating reference frame:
\beq
\label{momeq}
\frac{\partial^2}{\partial t^2} \bxi + 2 {\bf \Omega_s} \times \frac{\partial}{\partial t} \bxi  + \mathcal{C} \bxi = - \bnab U,
\eeq
where $U$ is the external gravitational potential, and $\mathcal{C} \bxi$ represents internal forces from the perturbed fluid. A free oscillation mode (indexed by the label $\alpha$) of the star satisfies
\beq
\label{modedef}
-\omega_\alpha^2 \bxi_\alpha - 2 i \omega_\alpha {\bf \Omega_s} \times \bxi_\alpha  + \mathcal{C} \bxi_\alpha = 0,
\eeq
for a mode with time dependence $\propto e^{- i \omega_\alpha t}$ (see \citealt{lai:06}). This convention implies that a mode with positive frequency and $m>0$ is prograde in the rotating frame of the star, while modes with positive frequency and $m<0$ are retrograde in the rotating frame (although they may be prograde in the inertial frame). 

In non-rotating stars, modes obey the orthonormality condition
\beq
\label{modeorth}
\langle \xi_\alpha | \xi_\beta \rangle \equiv \int dV \rho \bxi_\alpha^* \cdot \bxi_\beta = \delta_{\alpha \beta} \, ,
\eeq
where the integral extends over the volume of the star. However, in rotating stars (c.f. \citealt{schenk:02,lai:06}) modes obey the modified orthonomality condition
\beq
\label{modeorthrot}
\langle \xi_\alpha | \xi_\beta \rangle = \delta_{\alpha \beta} - \frac{2}{\omega_\alpha + \omega_\beta} W_{\alpha \beta},
\eeq
where
\beq
\label{Wdef}
W_{\alpha \beta} \equiv \int dV \rho \bxi_\alpha^* \cdot \Big( i {\bf \Omega_s} \times \bxi_\beta \Big) 
\eeq
is a Coriolis coupling coefficient. This modified orthonomality condition is important in accurately calculating the dynamical tidal response, especially when rotation strongly modifies the modes of interest. 

A general perturbation to the star must be expanded in terms of its displacement and velocity perturbations via a phase space mode expansion:
\beq
\label{expansion}
\left[ \begin{array}{c} \bxi \\ \partial \bxi / \partial t \end{array} \right] = \sum_\beta a_\beta \left[ \begin{array}{c} \bxi_\beta \\ - i \omega_\beta \bxi_\beta \end{array} \right],
\eeq
where $a_\beta$ is the time-dependent mode amplitude. The summation runs over both positive and negative values of the mode azimuthal number $m$, and also over both positive and negative eigenfrequencies $\omega_\beta$. Since the mode eigenfunction contains no time-dependence, equation \ref{expansion} implies that for any perturbation, 
\beq
\label{dmodedt}
\frac{\partial \bxi}{\partial t} = \sum_{\beta} \dot{a}_{\beta} \bxi_\beta = \sum_\beta -i \omega_\beta a_\beta \bxi_\beta,
\eeq
although individual terms in the summations are not generally equal to one another for a forced mode. Equation \ref{dmodedt} additionally leads to the relation 
\beq
\label{dmodedtbeta}
\sum_{\beta} \dot{a}_{\beta} \langle \xi_\alpha | \xi_\beta \rangle = \sum_\beta -i \omega_\beta a_\beta \langle \xi_\alpha | \xi_\beta \rangle,
\eeq

To obtain the tidal response of an oscillation mode (indexed by $\alpha$), we insert the mode expansion \ref{expansion} into equation \ref{momeq}, and take the inner product with $\bxi_\alpha$. Using the orthonormality condition \ref{modeorthrot} and the identity of equation \ref{dmodedtbeta}, one obtains the mode amplitude equation \citep{schenk:02}
\beq
\label{modeamp}
\dot{a}_\alpha + i \omega_\alpha a_\alpha = \frac{i}{2\omega_\alpha} \langle \xi_\alpha | - \bnab U \rangle.
\eeq
This equation describes the evolution of the mode amplitude $a_\alpha$, which includes both its equilibrium tide component and its dynamical tide component. Its precise form depends on the adopted normalization condition, which determines the meaning of the mode amplitude $a_\alpha$.

To understand tidal dynamics, it is useful to separate the equilibrium and dynamical tidal responses. To do this, we must subtract out the equilibrium tidal response from equation \ref{momeq}. The equilibrium tide is defined by taking the static ($\partial /\partial t \rightarrow 0$) limit of equation \ref{momeq}:
\beq
\label{eqtide}
\mathcal{C} \bxi_{\rm eq} \equiv - \bnab U.
\eeq
Equation \ref{eqtide} also implies that $\partial \bxi_{\rm eq} /\partial t = - i \omega \bxi_{\rm eq}$ for a tidal potential of form $U \propto e^{- i \omega t}$.

We then separate the dynamical and equilibrium tide components, $\bxi = \bxi_{\rm dyn} + \bxi_{\rm eq}$, which upon substitution into equation \ref{momeq} yields
\beq
\label{momeqdyn}
\frac{\partial^2}{\partial t^2} \bxi_{\rm dyn} + 2 {\bf \Omega_s} \times \frac{\partial}{\partial t} \bxi_{\rm dyn}  + \mathcal{C} \bxi_{\rm dyn} = - \frac{\partial^2}{\partial t^2} \bxi_{\rm eq} - 2 {\bf \Omega_s} \times \frac{\partial}{\partial t} \bxi_{\rm eq}.
\eeq
Next, we decompose the dynamical and equilibrium responses in the same fashion as equation \ref{expansion}:
\beq
\label{expansiondyn}
\left[ \begin{array}{c} \bxi_{\rm dyn} \\ \frac{\partial}{\partial t} \bxi_{\rm dyn} \end{array} \right] = \sum_\beta a_{\beta,{\rm dyn}} \left[ \begin{array}{c} \bxi_\beta \\ - i \omega_\beta \bxi_\beta \end{array} \right],
\eeq
and likewise for the equilibrium tide. Then, using the identity
\beq
\label{dmodedteq}
\sum_{\beta} \dot{a}_{\beta, {\rm eq}} \langle \xi_\alpha | \xi_\beta \rangle = -i \omega \sum_\beta a_{\beta, \rm eq} \langle \xi_\alpha | \xi_\beta \rangle = \sum_\beta -i \omega_\beta a_{\beta, \rm eq} \langle \xi_\alpha | \xi_\beta \rangle,
\eeq
and taking the same steps used to derive equation \ref{modeamp}, we find 
\beq
\label{modeampdyn1}
\dot{a}_{\alpha, {\rm dyn}} + i \omega_\alpha a_{\alpha,{\rm dyn}} = \frac{i}{2\omega_\alpha} \bigg[ 2 \omega \omega_\alpha a_{\alpha, {\rm eq}} - \sum_\beta \omega_\beta \omega_\alpha a_{\beta, {\rm eq}} \langle \xi_\alpha | \xi_\beta \rangle \bigg] .
\eeq
From equation \ref{eqtide} one can also find 
\beq
\label{modeampdyn2}
a_{\alpha, {\rm eq}} = \frac{1}{2\omega_\alpha^2} \langle \xi_\alpha | - \bnab U \rangle + \frac{1}{2\omega_\alpha^2} \sum_\beta a_{\beta, {\rm eq}} \omega_\beta \omega_\alpha \langle \xi_\alpha | \xi_\beta \rangle,
\eeq
and so
\beq
\label{modeampdyn3}
\dot{a}_{\alpha, {\rm dyn}} + i \omega_\alpha a_{\alpha,{\rm dyn}} = \frac{i \omega}{2\omega_\alpha^2} \langle \xi_\alpha | - \bnab U \rangle + \frac{i \omega}{2\omega_\alpha} (\omega - \omega_\alpha) \langle \xi_\alpha | \xi_{\rm eq} \rangle.
\eeq
Equation \ref{modeampdyn3} describes the evolution of the dynamical component of the mode amplitude $a_{\alpha,{\rm dyn}}$. It contains two forcing terms on the right-hand side, which we discuss below. 

The adiabatic solution to equation \ref{modeampdyn3} is obtained by assuming the magnitude of the mode amplitude changes slowly such that $\dot{a}_{\alpha, {\rm dyn}} = - i \omega a_{\alpha, {\rm dyn}}$. We allow for complex eigenfrequencies by replacing $\omega_\alpha$ with $\omega_\alpha + i \gamma_\alpha$, where $\gamma_\alpha$ is the mode growth rate which has been defined such that negative values of $\gamma_\alpha$ correspond to damped modes. The adiabatic mode amplitude is 
\beq
\label{modeampdyn}
a_{\alpha, {\rm dyn}} = \frac{1}{2} \frac{\omega}{\omega_\alpha - \omega + i \gamma_\alpha} \frac{\langle \xi_\alpha | - \bnab U \rangle}{\omega_\alpha^2} - \frac{1}{2} \frac{\omega}{\omega_\alpha + i \gamma_\alpha} \langle \xi_\alpha | \xi_{\rm eq} \rangle.
\eeq
Here, we have assumed for simplicity that the mode is weakly damped, i.e., $\gamma_\alpha \ll \omega_\alpha$, which is always the case for detectable modes in stars.

Equation \ref{modeampdyn} describes the dynamical response of a mode $\alpha$ to tidal forcing. The first term is the same as the total mode amplitudes discussed in previous works, but multiplied by a factor $\omega/\omega_\alpha$. This term is largest near resonances where $\omega \simeq \omega_\alpha$.  The multiplicative factor $\omega/\omega_\alpha$ does not strongly affect nearly resonant modes, but is important for cancellations between non-resonant modes. As we show below, these cancellations entail that high frequency (relative to the tidal forcing frequency) f-modes which dominate the equilibrium tide response have little contribution to the dynamical tide.

The second term in equation \ref{modeampdyn} is due to damping of the equilibrium tide. Upon summation over both a mode and its complex conjugate, it produces a term
\beq
a_\alpha + a_\alpha^* = \frac{i \omega \gamma_\alpha}{\omega_\alpha^2 + \gamma_\alpha^2} \langle \xi_\alpha | \xi_{\rm eq} \rangle.
\eeq
This term accounts for the lag of the equilibrium tide due to damping, generating a phase shift in the equilibrium tidal bulge. The summation of this term over all modes accounts for the classical tidal lag angle $\delta \propto 1/Q$ associated with the equilibrium tide, where $Q$ is the equilibrium tide quality factor.

Finally, recall that equation \ref{modeampdyn} is only the {\it forced} mode amplitude. The free oscillation modes of the star satisfy 
\beq
\label{modeampfree}
\dot{a}_{\alpha, {\rm free}} + i \omega_\alpha a_{\alpha,{\rm free}} = 0.
\eeq
Thus, the stellar oscillation modes are still permitted to oscillate at their natural oscillation frequency, $\omega_\alpha$, at an arbitrary amplitude. Therefore, stellar oscillation modes of stars in close binaries have three simultaneous amplitudes: $a_{\rm free}$, $a_{\rm eq}$, and $a_{\rm dyn}$.

Using similar notation as \cite{fullerkoi54:12}, the dynamical part of the mode amplitude (computed from the first term of equation \ref{modeampdyn}, in the rotating frame of the star for aligned spin and orbit) is then
\beq
\label{modedyn}
a_{\alpha,{\rm dyn},N} = \frac{1}{2} \epsilon_{lm} F_{Nm} Q_\alpha \frac{\omega_{Nm}}{\omega_\alpha - \omega_{Nm} + i \gamma_\alpha} e^{-i \omega_{Nm} t}
\eeq
where 
\beq
\label{epslm}
\epsilon_{lm} = \epsilon_l \, W_{lm}
\eeq
describes the strength of the tidal forcing due to the component of the tidal potential with indices $l$ and $m$, $W_{22} = W_{2-2} = \sqrt{3 \pi/10}$, and $W_{20} = -\sqrt{\pi/5}$. We generalize to the misaligned case in Section \ref{misalignment}.

The formalism used above is strictly only valid for adiabatic modes, as non-adiabatic modes do not obey the same orthogonality conditions. We suspect non-adiabatic corrections to equation \ref{modeampdyn3} will be of order $\gamma_\alpha a_\alpha$. For modes far from resonance, we expect these corrections to be negligible. However, for highly damped modes with $\gamma_\alpha \sim \Delta \omega_\alpha$, i.e., modes near the traveling wave regime, the corrections may be significant. There, the physical picture is a train of tidally excited gravity waves propagating into the stellar envelope, as examined for early type stars \citep{zahn:75,zahn:77} and white dwarfs \citep{fullerwd:12,burkart:13}. In this case, there are no resonances in the tidal response, and the wave amplitude is a smoothly varying function of frequency. This situation is similar to the result of equation \ref{modeampdyn}, because the amplitude of the resonant detuning term  is smoothly varying and can only vary by a factor of $\sim 2$ when $\gamma_\alpha \sim \Delta \omega_\alpha$. Hence, in this circumstance, we suspect our formalism still yields approximately correct results (at the factor of 2 level), but it cannot be used for precise predictions of TEO amplitudes or phases.

\section{Summation over Oscillation Modes}
\label{sum}

The forced response at a given orbital harmonic $N$ is found by summing over all modes:
\beq
\label{xin}
\bxi_{N,{\rm dyn}} = \sum_{\alpha} a_{\alpha,{\rm dyn},N} \bxi_\alpha
\eeq
This sum over $\alpha$ includes all modes which significantly couple with the tidal potential. For a given value of $m$, this includes modes with both positive and negative frequencies $\omega_\alpha$. Note that a mode with $(\omega_\alpha,m)$ has a physically identical counterpart with $(-\omega_\alpha,-m)$, and both must be formally included in the summation. We will refer to the former mode as the $\omega_{\alpha,m}$ mode and the latter as the $\omega_{-\alpha,-m}$ mode. We must also include forcing at both $N$ and $-N$ since the decomposition of the tidal potential into orbital harmonics includes both positive and negative forcing frequencies. The observed response at a harmonic $|N|$ is a combination of both the positive and negative forcing frequencies. In what follows, we will consider the $m=0$ and $m=\pm2$ components of the $l=2$ tidal potential.\footnote{As is commonly done, we ignore components of the tidal potential with $l>2$ because the strength of the potential scales as $(R/a)^{(l+1)}$.} We only compute the contributions from $m=0$ and $m=\pm2$ modes because the $m=\pm1$ contributions vanish if the spin and orbit are aligned. However, for misaligned spins, the $m=\pm1$ terms should be included. 

Including the contributions listed above, the tidal response at a harmonic $|N|$ of the orbital frequency is, in the inertial frame, for $|m|>0$ modes
\begin{align}
\label{xin2}
\bxi_{|N|,{\rm dyn}} = \sum_{\omega_\alpha > 0, m>0} \frac{\epsilon_{lm}}{2} \Bigg[& Q_{\alpha m} F_{Nm} H_{\alpha m}(\theta) \bxi_{\alpha,m}(r) \frac{\omega_{Nm}}{\omega_{\alpha m} - \omega_{Nm} + i \gamma_{\alpha m} } e^{-i(N \Omega t - m \phi)}  \nonumber \\
&+ Q_{-\alpha -m} F_{-N-m} H_{-\alpha -m}(\theta) \bxi_{-\alpha,-m}(r)  \frac{\omega_{-N-m}}{\omega_{-\alpha -m} - \omega_{-N-m} + i \gamma_{-\alpha -m} } e^{i(N \Omega t - m \phi)} \nonumber \\
&+ Q_{-\alpha m} F_{Nm} H_{-\alpha m(\theta)} \bxi_{-\alpha,m}(r)  \frac{\omega_{Nm}}{\omega_{-\alpha m} - \omega_{Nm} + i \gamma_{-\alpha m} } e^{-i(N \Omega t - m \phi)} \nonumber \\
&+ Q_{\alpha -m} F_{-N-m} H_{\alpha -m}(\theta) \bxi_{\alpha,-m}(r)  \frac{\omega_{-N-m}}{\omega_{\alpha -m} - \omega_{-N-m} + i \gamma_{\alpha -m} } e^{i(N \Omega t - m \phi)} \nonumber \\
&+ {\rm Retrograde \ Terms} \Bigg].
\end{align}
Here, we have factored out the angular dependence of the eigenfunctions in the traditional approximation, $\bxi_\alpha = \bxi_\alpha(r) H_{\alpha m}(\theta) e^{i m \phi}$, where $H_{\alpha m}(\theta)$ is the Hough function associated with a mode $\alpha$ (see Section \ref{trad}). We have also omitted the retrograde terms (those with dependence $\propto e^{\pm i (N \Omega t + m \phi)}$) because they are typically excited to very low amplitudes since the value of $F_{N-m}=F_{-Nm}$ is very small for any realistic value of the orbital eccentricity. Here, retrograde signifies modes that are retrograde in the {\it inertial} frame, i.e., modes whose patterns revolve in the opposite direction of the orbital motion. We may still obtain large amplitude modes that are prograde in the inertial frame, yet are retrograde in rotating frame of the star.

To simplify equation \ref{xin2}, we use the fact that $\omega_{Nm} = - \omega_{-N-m}$, $\omega_{\alpha m} = - \omega_{-\alpha -m}$, $F_{Nm} = F_{-N-m}$, $Q_{\alpha m} = Q_{-\alpha -m}$, $H_{\alpha m}(\theta) = H_{-\alpha -m}(\theta)$, $\bxi_{\alpha m}(r) = \bxi_{-\alpha -m}(r)$, $\gamma_{\alpha m} = \gamma_{-\alpha -m}$, and likewise for $(-\alpha, m)$ and $(\alpha, -m)$ combinations. After careful manipulation we obtain
\begin{align}
\label{xinm2}
\bxi_{|N|,{\rm dyn}} = \sum_{\omega_\alpha > 0, m>0} & \epsilon_{lm} Q_{\alpha m} F_{Nm} H_{\alpha m}(\theta) \bxi_{\alpha m}(r) \frac{\omega_{Nm}}{\sqrt{(\omega_{\alpha m} - \omega_{Nm})^2 + \gamma_{\alpha m}^2 }} \sin \big(N \Omega t - m \phi + \delta_{\alpha m} \big) \nonumber \\
&+ \epsilon_{lm} Q_{\alpha -m} F_{Nm} H_{\alpha -m}(\theta) \bxi_{\alpha -m}(r) \frac{\omega_{Nm}}{\sqrt{(\omega_{\alpha -m} + \omega_{Nm})^2 + \gamma_{\alpha -m}^2 }} \sin \big(N \Omega t - m \phi + \delta_{\alpha -m} \big) \nonumber \\
&+ {\rm Retrograde \ Terms},
\end{align}
with
\beq
\delta_{\alpha m} = {\rm atan2} \bigg[\frac{\omega_{\alpha m} - \omega_{Nm}}{-\gamma_{\alpha m}} \bigg]
\eeq
and
\beq
\label{xidm2b}
\delta_{\alpha -m} = {\rm atan2} \bigg[ \frac{-\omega_{\alpha -m} -\omega_{Nm}}{-\gamma_{\alpha -m}} \bigg].
\eeq
Here, the ${\rm atan2}$ function must be used to obtain the correct phase shift. These equations assume that the mode displacement $\bxi$ is a purely real quantity, i.e., there are no non-adiabatic effects. We examine non-adiabatic effects in more detail in Section \ref{lum}.

The first line in equation \ref{xinm2} represents prograde modes forced in the prograde direction, while the second term represents retrograde modes forced in the prograde direction. In some previous works, the second line has been omitted. However, it is important to retain it for two reasons. First, it can still to lead to resonant excitation because the forcing frequency $\omega_{Nm} = N \Omega - m \Omega_s$ is negative for $N \Omega < m \Omega_s$. Physically this corresponds to retrograde modes (in the rotating frame) being forced in the prograde direction (in the inertial frame). Second, terms on the first and second lines tend to cancel each other (due to their nearly opposite phase shifts) for high frequency modes with $\omega_{\alpha m}, \omega_{\alpha -m} \gg \omega_{Nm}$. Consequently the stellar f modes that dominate the equilibrium tide deconstructively interfere such that their contribution to the dynamical tide is substantially reduced.

The same approach can be used for $m=0$ modes. In this case there are no terms corresponding to $(\alpha, -m)$ combinations, but the $(-N, m)$ combinations  cannot be dropped because $F_{-Nm}=F_{Nm}$ for $m=0$. The result is
\begin{align}
\label{xinm0}
\bxi_{|N|,{\rm dyn}} = \sum_{\omega_\alpha > 0, m=0} & \epsilon_{lm} Q_{\alpha m} F_{Nm} H_{\alpha m}(\theta) \bxi_{\alpha,m}(r) \frac{\omega_{Nm}}{\sqrt{(\omega_{\alpha m} - \omega_{Nm})^2 + \gamma_{\alpha m}^2 }} \sin \big(N \Omega t + \delta_{\alpha m} \big) \nonumber \\
&+ \epsilon_{lm} Q_{\alpha m} F_{Nm} H_{\alpha m}(\theta) \bxi_{\alpha m}(r) \frac{\omega_{Nm}}{\sqrt{(\omega_{\alpha m} + \omega_{Nm})^2 + \gamma_{\alpha m}^2 }} \sin \big(N \Omega t + \delta_{-\alpha m} \big)
\end{align}
with
\beq
\delta_{\alpha m} = {\rm atan2} \bigg[ \frac{\omega_{\alpha m}-\omega_{Nm}}{-\gamma_{\alpha m}} \bigg]
\eeq
and
\beq
\label{xidm0b}
\delta_{-\alpha m} = {\rm atan2} \bigg[ \frac{-\omega_{\alpha m} - \omega_{Nm}}{-\gamma_{\alpha m}} \bigg].
\eeq
Once again, the second term in equation \ref{xinm0} has been dropped in some previous works. In this case, it cannot lead to resonant excitation, although it must still be retained in order to obtain the deconstructive interference described above. Note that the form of equations \ref{xinm0}-\ref{xidm0b} is the same as equations \ref{xinm2}-\ref{xidm2b} without the retrograde terms, hence the latter can be used for $m=0$ modes.

\subsection{Luminosity Fluctuations and Non-adiabatic Effects}
\label{lum}

TEOs can be observed with high-accuracy photometric data, for which the observable feature of TEOs is a disk-integrated luminosity fluctuation $\Delta L/L$. Both the amplitude and phase of these oscillations can be calculated, so a single eccentric binary can yield large numbers of observable quantities (the orbital harmonics, amplitudes, and phases of TEOs) which may be compared with tidal theories. 

A complication not discussed above is that TEOs may become strongly non-adiabatic near the surface of the star where they are observed. The mode eigenfunction $\bxi(r)$ will thus be phase-shifted relative to an adiabatic mode. This is equivalent to the modes obtaining imaginary components to their eigenfunctions. Then any mode quantity, e.g., the radial component of the displacement $\xi_r$, may be expressed as 
\beq
\label{imreal}
\xi_{r,\alpha}(r) = \xi^{\rm R}_{r,\alpha}(r) + i \xi^{\rm I}_{r,\alpha}(r).
\eeq
The mode overlap integral $Q_\alpha$ also obtains an imaginary component. Recomputing the sums above, we find
\begin{align}
\label{xinm2i}
\xi_{r,|N|,{\rm dyn}} = \sum_{\omega_\alpha > 0, m \geq 0} & \epsilon_{lm} F_{Nm} H_{\alpha m}(\theta) \big| Q_{\alpha m} \xi_{r,\alpha m}(r) \big| \frac{\omega_{Nm}}{\sqrt{(\omega_{\alpha m} - \omega_{Nm})^2 + \gamma_{\alpha m}^2 }} \sin \big(N \Omega t - m \phi + \delta_{\alpha m} \big) \nonumber \\
&+ \epsilon_{lm}  F_{Nm} H_{\alpha -m}(\theta) \big| Q_{\alpha -m} \xi_{r,\alpha -m}(r) \big| \frac{\omega_{Nm}}{\sqrt{(\omega_{\alpha -m} + \omega_{Nm})^2 + \gamma_{\alpha -m}^2 }} \sin \big(N \Omega t - m \phi + \delta_{\alpha -m} \big) \nonumber \\
&+ {\rm Retrograde \ Terms},
\end{align}
with 
\beq
\label{phasem2}
\delta_{\alpha m} = {\rm atan2} \bigg[ \frac{(\omega_{\alpha m} - \omega_{Nm}) (Q_{\alpha m}\xi_{r,\alpha m})^{\rm R} + \gamma_{\alpha m} (Q_{\alpha m}\xi_{r,\alpha m})^{\rm I}}{(\omega_{\alpha m} - \omega_{Nm}) (Q_{\alpha m}\xi_{r,\alpha m})^{\rm I} - \gamma_{\alpha m} (Q_{\alpha m}\xi_{r,\alpha m})^{\rm R}} \bigg]
\eeq
and
\beq
\label{phasem2n}
\delta_{\alpha -m} = {\rm atan2} \bigg[ \frac{-(\omega_{\alpha -m} + \omega_{Nm})(Q_{\alpha -m}\xi_{r,\alpha -m})^{\rm R} + \gamma_{\alpha -m} (Q_{\alpha -m}\xi_{r,\alpha -m})^{\rm I}}{-(\omega_{\alpha -m} + \omega_{Nm}) (Q_{\alpha -m}\xi_{r,\alpha -m})^{\rm I} - \gamma_{\alpha -m} (Q_{\alpha -m}\xi_{r,\alpha -m})^{\rm R}} \bigg] \, .
\eeq
Here, $\big| Q_{\alpha m} \xi_{r,\alpha m}(r) \big| = \big[ {(Q_{\alpha m}\xi_{r,\alpha m})^{\rm R}}^2 + {(Q_{\alpha m}\xi_{r,\alpha m})^{\rm I}}^2 \big]^{1/2}$, and $(Q_{\alpha m}\xi_{r,\alpha m})^{\rm R} = Q_{\alpha m}^{\rm R} \xi_{r,\alpha m}^{\rm R} - Q_{\alpha m}^{\rm I} \xi_{r,\alpha m}^{\rm I}$, and $(Q_{\alpha m}\xi_{r,\alpha m})^{\rm I} = Q_{\alpha m}^{\rm R} \xi_{r,\alpha m}^{\rm I} + Q_{\alpha m}^{\rm I} \xi_{r,\alpha m}^{\rm R}$. A similar equation holds for any perturbed quantity, e.g., the perturbed flux $\Delta F/F$, but with $\xi_r$ replaced by $\Delta F/F$ at each point in equations \ref{xinm2i}-\ref{phasem2n}. These equations also apply for $m=0$ modes, except for the additional retrograde terms.

\subsubsection{Observed Luminosity Fluctuation}

The observable luminosity fluctuations are disk integrated quantities containing three contributing effects: surface area perturbations, surface normal perturbations, and flux perturbations. These have been examined in \cite{buta:79,robinson:82,townsend:03}, but here we follow the procedure of \cite{burkart:12}. For a luminosity fluctuation due to a single mode with spherical harmonic angular dependence, the magnitude of the luminosity fluctuation is
\beq
\label{dlum}
\frac{\Delta L_\alpha}{L} = \bigg[ (2 b_l - c_l) \frac{\xi_{r, \alpha}(R)}{R} + b_l \frac{\Delta F_\alpha(R)}{F(R)} \bigg] Y_{lm}(i_s,\phi_s),
\eeq
where $b_l$ and $c_l$ are limb darkening coefficients and $i_s$ and $\phi_s$ are the angular coordinates of the observer in the star's rotating frame at $t=0$. Note that $\phi_s$ is related to the argument of periapsis $\omega_{\rm p}$ of the secondary via
\beq
\label{phio}
\phi_s = -\pi/2 - \omega_{\rm p} \, .
\eeq
From radial velocity measurements, one typically measures the argument of periapsis of the orbit of the primary, which is shifted by $\pi$ relative to $\omega_{\rm p}$ of the secondary. 

The quantity $\Delta F_\alpha$ is the perturbed flux. In rotating stars, modes have Hough function angular dependence and so the result will change somewhat. However, we note that any Hough function may always be decomposed in terms of spherical  harmonics,
\beq
\label{hough}
H_{km}(\theta,\phi) = \sum_{l} h_{klm} Y_{lm}(\theta,\phi)
\eeq
where $h_{klm}$ is the angular overlap integral of equation \ref{hdef}. Because high $l$ spherical harmonics suffer large cancelation effects when integrated over the stellar disk, only the low $l$ components of a Hough function will contribute appreciably to the disk-averaged luminosity fluctuation of a mode. For TEOs, the $l=2$ components will dominate. Therefore, to good approximation, the luminosity fluctuation of a TEO in the traditional approximation is 
\beq
\label{dlum2}
\frac{\Delta L_\alpha}{L} = \bigg[ (2 b_2 - c_2) \frac{\xi_{r, \alpha}(R)}{R} + b_2 \frac{\Delta F_\alpha(R)}{F(R)} \bigg] h_{k2m} Y_{2m}(i_s,\phi_s).
\eeq

Using equation \ref{dlum2}, we find that the net luminosity fluctuation at a harmonic $N$ of the orbital frequency is
\begin{align}
\label{dlum3}
\frac{\Delta L_{|N|,{\rm dyn}}}{L} \simeq \sum_{\omega_\alpha > 0, m \geq 0} & \epsilon_{lm} F_{Nm} Y_{2m}(i_s,0) \big| Q_{\alpha m} L_{\alpha m} \big| \frac{\omega_{Nm}}{\sqrt{(\omega_{\alpha m} - \omega_{Nm})^2 + \gamma_{\alpha m}^2 }} \sin \big(N \Omega t - m \phi_s + \delta_{\alpha m} \big) \nonumber \\
&+ \epsilon_{lm} F_{Nm} Y_{2m}(i_s,0) \big| Q_{\alpha -m} L_{\alpha -m} \big| \frac{\omega_{Nm}}{\sqrt{(\omega_{\alpha -m} + \omega_{Nm})^2 + \gamma_{\alpha -m}^2 }} \sin \big(N \Omega t - m \phi_s + \delta_{\alpha -m} \big) \nonumber \\
&+ {\rm Retrograde \ Terms},
\end{align}
with
\beq
\label{phaselum}
\delta_{\alpha m} = {\rm atan2} \bigg[ \frac{(\omega_{\alpha m} - \omega_{Nm}) (Q_{\alpha m} L_{\alpha m})^{\rm R} + \gamma_{\alpha m} (Q_{\alpha m} L_{\alpha m})^{\rm I} }{(\omega_{\alpha m} - \omega_{Nm}) (Q_{\alpha m} L_{\alpha m})^{\rm I} - \gamma_{\alpha m} (Q_{\alpha m} L_{\alpha m})^{\rm R} } \bigg]
\eeq
and
\beq
\label{phaselum2}
\delta_{\alpha -m} = {\rm atan2} \bigg[ \frac{-(\omega_{\alpha -m} + \omega_{Nm}) (Q_{\alpha -m} L_{\alpha -m})^{\rm R} + \gamma_{\alpha -m} (Q_{\alpha -m} L_{\alpha -m})^{\rm I} }{-(\omega_{\alpha -m} + \omega_{Nm}) (Q_{\alpha -m} L_{\alpha -m})^{\rm I} - \gamma_{\alpha -m} (Q_{\alpha -m} L_{\alpha -m})^{\rm R} } \bigg],
\eeq
and
\beq
\label{dl}
L_{\alpha m} = (2 b_2 - c_2) h_{k2m} \frac{\xi_{r, \alpha m}(R)}{R} + b_2 h_{k2m} \frac{\Delta F_{\alpha m}(R)}{F(R)} .
\eeq

The sum of sinusoidal oscillations in equation \ref{dlum3} may be written as a single sinusoidal oscillation of form
\beq
\label{dlumtot}
\frac{\Delta L_{|N|,{\rm dyn}}}{L} = A_N \sin(N \Omega t + \delta_N) \, ,
\eeq
where the amplitude of the oscillation is
\beq
\label{AN}
A_N = \bigg[ \sum_{i,j} A_i A_j \cos(\Delta_i - \Delta_j) \bigg]^{1/2}
\eeq
and the amplitudes $A_i$ are given by the factors in front of the $\sin$ terms in equation \ref{dlum3}, i.e., 
\beq
\label{Ai}
A_i = \epsilon_{lm} F_{Nm} Y_{2m}(i_s,0) \big| Q_{\alpha m} L_{\alpha m} \big| \frac{\omega_{Nm}}{\sqrt{(\omega_{\alpha m} - \omega_{Nm})^2 + \gamma_{\alpha m}^2 }} \, .
\eeq
The phases $\Delta_i$ are given by the phases of the $\sin$ terms in equation \ref{dlum3},
\beq
\label{deltai}
\Delta_i = \delta_{\alpha m} - m \phi_s \, .
\eeq
Additionally, the phase $\delta_N$ of the observed oscillation is 
\beq
\label{deltaN}
\delta_N = {\rm atan2} \bigg[ \frac{\sum A_i \sin \Delta_i}{\sum A_i \cos \Delta_i} \bigg] \, .
\eeq
The sums over the indices $i$ and $j$ in equations \ref{AN} and \ref{deltaN} must run over all terms that contribute to the luminosity fluctuation at orbital harmonic $N$. This includes terms in equation \ref{dlum3} for both $m=2$ and $m=0$ modes, which may interfere with another. In systems where both stars are pulsating, contributions from both stars must be included.  In many cases, however, one star may dominate the pulsation signal, and either the $m=0$ or $|m|=2$ terms may dominate, and we may limit the sums to that value of $m$.\footnote{Of course, the $l>2$ components of the tidal potential also contribute, as do additional branches of Hough functions (corresponding to $l>2$ modes in the non-rotating limit). However, as discussed above, these contributions are small in many circumstances.}

The frequency $N \Omega$, luminosity fluctuation $A_N$, and phase $\delta_N$ from equation \ref{dlumtot} represent the observable quantities for TEOs in photometric data. Recall that our expansion of the time dependence of the tidal potential into orbital harmonics requires that $t=0$ corresponds to the time of periastron. Therefore, reliably measuring mode phases from photometric data requires a determination of both the epoch of periastron and the argument of periapsis. As discussed in Section \ref{gmodes}, the value of $L_\alpha$ generally has a substantial imaginary component, whereas $\gamma_\alpha$ is typically much smaller than the resonant detuning, unless the mode is extremely close to resonance. Consequently, the observed phase shift will almost always be dominated by non-adiabatic effects that contribute to the imaginary part of $L_\alpha$.

It is apparent from the tedious summations above that the relation between observed luminosity fluctuations and stellar properties is not transparent. The values $Q_\alpha$, $L_\alpha$, $\omega_\alpha$, and $\gamma_\alpha$ must be calculated for the modes of a given stellar model, and can be somewhat sensitive to the stellar parameters. Furthermore, the values of $\epsilon_{lm}$, $F_{Nm}$, $Y_{lm}(i_s,0)$, $\omega_{Nm}$ require good measurements of the orbital parameters (and stellar masses, radii, and spins) to be accurately calculated. Most important, the resonant detuning term $\omega_{Nm}/\sqrt{(\omega_{\alpha m} - \omega_{Nm})^2 + \gamma_{\alpha m}^2 }$ is extremely sensitive to the precise values of the mode frequencies $\omega_\alpha$. For these reasons, reliably performing tidal asteroseismology (i.e., using the mode frequencies, amplitudes, and phases to constrain stellar parameters) is difficult in practice.

\section{Accounting for Spin-Orbit Misalignment}
\label{misalignment}

Here we expand the formalism above to allow for misalignment between the stellar spin axis and the orbital angular momentum axis. This has been discussed briefly in \cite{lai:97}, here we revisit this calculation to predict amplitudes and phases of tidally excited modes in misaligned binaries. 

We begin from the amplitude equation \ref{modeampdyn3} which applies in the rotating frame of the star with $\hat{z}_{\rm spin}$ in the direction of the spin angular momentum,
\beq
\label{amp1}
\dot{a}_\alpha + i \omega_\alpha a_\alpha = \frac{i \omega_{Nm}}{2\omega_\alpha^2} \langle \xi_\alpha | - \bnab U \rangle \ .
\eeq
We have dropped the ${\rm dyn}$ subscript for simplicity. Performing the standard decomposition of the tidal potential $U$ in the rotating frame gives
\beq
\label{U1}
U = - G M' \sum_l \frac{4 \pi}{2 l + 1} \frac{r^l}{D'^{l+1}} \sum_m Y_{lm}(\theta,\phi) Y_{l-m}(\theta',\phi') \, .
\eeq
Here, ($D'$,$\theta'$,$\phi'$) is the time-dependent location of the secondary star in the rotating frame of the primary. Using $Y_{l-m} = (-1)^m Y_{lm}^*$, each component of the tidal potential has form
\beq
\label{U2}
U_{lm} = - G M' \frac{4 \pi}{2 l + 1} \frac{r^l}{D'^{l+1}} (-1)^m Y_{lm}(\theta,\phi)  Y_{lm}^*(\theta',\phi') \, .
\eeq

To express quantities in reference frames rotated relative to one another, we expand in Wigner $\mathcal{D}$ functions,
\beq
\label{wignerD}
Y_{lm}(\theta',\phi') = \sum_{m_o=-l}^{l} \mathcal{D}^l_{m_o m} (\alpha,\beta,\gamma) Y_{l m_o} (\theta'_o,\phi'_o) \, .
\eeq
Here, $\alpha$, $\beta$, and $\gamma$ are Euler angles describing the rotations necessary to transform the new orbital frame to the rotating frame, and $\theta'_o$,$\phi'_o$ are the location of the secondary in the new reference frame. We  define the orbital frame as having $\hat{z}_{\rm orb}$ in the direction of the orbital angular momentum vector, and $\hat{x}_{\rm  orb}$ in the direction of the secondary at periastron. The value of the Wigner function is 
\beq
\label{wignerD2}
\mathcal{D}^l_{m_o m} (\alpha,\beta,\gamma) = e^{-i m_o \alpha} d^l_{m_o m}(\beta) e^{-i m \gamma} \, ,
\eeq
where $d^l_{m_o m}(\beta)$ is an element of Wigner's small d-matrix. We caution that the definitions of Wigner functions and Euler angles is dependent on the chosen convention for the series of rotations to transform between coordinate systems. Equation \ref{wignerD2} uses the quantum mechanics convention used in \cite{varshalovich:88} and adopts a z-y-z convention for the series of rotations $\alpha,\beta,\gamma$, which differ from the classical Euler angles used below that follow a z-x-z convention.

\begin{figure*}
\begin{center}
\includegraphics[scale=0.6]{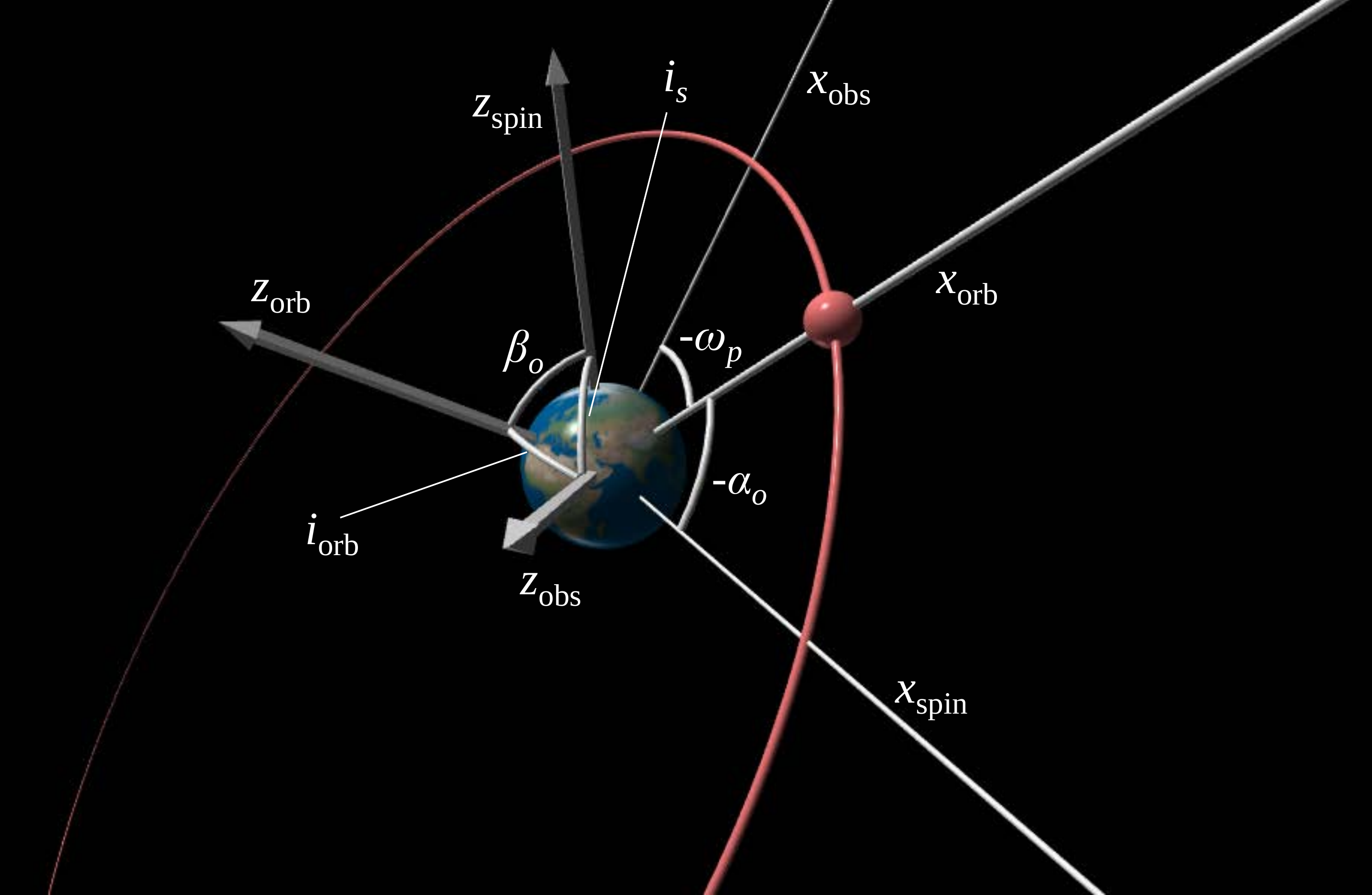}
\end{center} 
\caption{ \label{Geometry} Three-dimensional representation of an eccentric orbit misaligned with the spin of the primary star, which is given an Earth texture to clarify its spin axis. The secondary star is shown by a red sphere, and the diagram corresponds to time $t=0$ when the secondary is at periastron. The $x$ and $z$-axes of the three coordinate frames discussed in the text are labeled. All $x$-axes lie in the orbital plane at $t=0$.}
\end{figure*}

To transform the orbital frame to the rotating frame, we can rotate it about the $z_{\rm orb}$ axis by an angle $\alpha_o$, then around its new $x-$axis by an angle $\beta_o$, then about its new $z$-axis by an angle $\gamma_o$. We define the $\hat{x}_{\rm spin}$ of the rotating frame such that it lies in the orbital plane at time $t=0$. The angle $\alpha_o$ is the angle between $\hat{x}_{\rm spin}$ and $\hat{x}_{\rm orb}$ at $t=0$, and is the longitudinal spin-orbit misalignment measured in the orbital plane. The angle $\beta_o$ is the misalignment angle between $\hat{z}_{\rm spin}$ and $\hat{z}_{\rm orb}$, and our definition of $\hat{x}_{\rm spin}$ entails $\gamma_o = \Omega_s t$, i.e., it increases with time due to the rotation of the star. Note that this series of rotations uses a z-x-z sequence corresponding to classical Euler angles that is different from the quantum convention used in the $\mathcal{D}$ functions above. The two are equivalent if we use $\alpha = \alpha_o - \pi/2$, $\beta = \beta_o$, and $\gamma = \gamma_o + \pi/2$. Figure \ref{Geometry} shows a diagram of our reference frames.

The tidal potential in the rotating frame is thus
\beq
\label{U3}
U_{lm} = - G M' \frac{4 \pi}{2 l + 1} \frac{r^l}{D'^{l+1}} (-1)^m e^{i m (\Omega_s t + \pi/2)} Y_{lm}(\theta,\phi) \sum_{m_o} e^{i m_o(\alpha_o - \pi/2)} d^l_{m_o m}(\beta_o) \, Y^*_{l m_o} (\theta'_o,\phi'_o) \, .
\eeq
In the orbital frame, the secondary lies in the equator at $\theta'_o=\pi/2$ such that only even values of $m_o$ contribute to the sum. Inserting this expression into equation \ref{amp1}, we obtain 
\beq
\label{amp2}
\dot{a}_\alpha + i \omega_\alpha a_\alpha = \frac{i \omega_{Nm}}{2} \epsilon_l Q_{\alpha m} (-1)^m e^{i m (\Omega_s t + \pi/2)} \sum_{{\rm even} \, m_o} \bigg( \frac{a}{D'} \bigg)^{l+1} W_{l m_o} d^l_{m_o m}(\beta_o) e^{-i m_o (\phi'_o - \alpha_o + \pi/2)} \, .
\eeq
As above, we can decompose the forcing terms into orbital harmonics to obtain the forced mode amplitude at each orbital harmonic $N$
\beq
\label{amp3}
\dot{a}_{\alpha,N} + i \omega_\alpha a_{\alpha,N} = \frac{i \omega_{Nm}}{2} \epsilon_l Q_{\alpha m} (-1)^m e^{i m \pi/2} e^{-i \omega_{N m} t} \sum_{{\rm even} \, m_o} W_{l m_o} d^l_{m_o m}(\beta_o) F_{N m_o} e^{i m_o ( \alpha_o - \pi/2)}   \, ,
\eeq
with $F_{Nm}$ defined in equation \ref{FN}. 

Then, as above, the forced amplitude of each mode at each orbital harmonic $N$ is  
\beq
\label{amp4}
a_{\alpha,N} = \frac{1}{2} \epsilon_l Q_{\alpha m}  e^{i m \pi/2} \frac{ \omega_{Nm}}{\omega_\alpha - \omega_{Nm} + i \gamma_\alpha}  e^{-i \omega_{N m} t} \sum_{{\rm even} \, m_o} (-1)^m W_{l m_o} d^l_{m_o m}(\beta_o) F_{N m_o} e^{i m_o ( \alpha_o - \pi/2)}   \, ,
\eeq
For arbitrary spin-orbit misalignment, odd $m$ terms can be excited, in contrast the aligned case. The misalignment angle $\beta_o$ modifies the strength of the forcing for different values of $m$, while the angle $\alpha_o$ induces a phase shift. 

To simplify equation \ref{amp2}, we can combine the terms in the sum into a single forcing term:
\beq
\label{Xdef}
\sum_{{\rm even} \, m_o} (-1)^m  W_{l m_o} d^l_{m_o m}(\beta_o) F_{N m_o} e^{i m_o ( \alpha_o - \pi/2)} = X_{Nm} e^{ix_{Nm}} \, 
\eeq
with amplitude $X_{Nm}$ of 
\beq
\label{Xnm}
X_{Nm} = \bigg[ \bigg( \sum_{m_o} W_{l m_o} d^l_{m_o m} F_{N m_o} \cos \big[m_o (\alpha_o - \pi/2) + m \pi \big] \bigg)^{\! 2} + \bigg( \sum_{m_o} W_{l m_o} d^l_{m_o m} F_{N m_o} \sin \big[m_o (\alpha_o - \pi/2) + m \pi \big] \bigg)^{\! 2} \bigg]^{1/2}
\eeq
and phase shift 
\beq
\label{x}
x_{Nm} = {\rm atan2} \bigg[ \sum_{m_o} W_{l m_o} d^l_{m_o m} F_{N m_o} \sin \big[m_o (\alpha_o - \pi/2) + m \pi \big] , \sum_{m_o} W_{l m_o} d^l_{m_o m} F_{N m_o} \cos \big[m_o (\alpha_o - \pi/2) + m \pi \big]\bigg] \, .
\eeq
In the limit of aligned spin and orbit, $\beta_o \rightarrow 0$ and we can choose $\alpha_o=0$. Then $d^l_{m_o m}(\beta_o) \rightarrow \delta_{m_o m}$, $X_{Nm} \rightarrow W_{lm} F_{Nm}$, $x_{Nm} \rightarrow m \pi/2$, and equation \ref{amp4} reduces to equation \ref{modedyn}.

\subsection{Observed Pulsation Amplitude and Phase}

At any instant, the observed luminosity variation due to a mode forced at orbital harmonic $N$ is
\beq
\label{dlumN}
\frac{\Delta L_{\alpha,N}}{L} = a_{\alpha,N} L_\alpha Y_{2m}(i_s,\phi_s) \, .
\eeq
To obtain the observed luminosity fluctuation we must calculate $Y_{2m}(i_s,\phi_s)$. To do this, we can first transform back into the orbital frame, 
\begin{align}
\label{wignerD3}
Y_{lm}(i_s,\phi_s) &= \sum_{m_o=-l}^{l} \mathcal{D}^l_{m_o m} (\alpha,\beta,\gamma) Y_{l m_o} (\theta_{\rm obs},\phi_{\rm obs}) \nonumber \\
&= \sum_{m_o=-l}^{l} e^{-i m_o (\alpha_o - \pi/2)} d^l_{m_o m} (\beta_o) e^{- i m (\Omega_s t + \pi/2)} Y_{l m_o} (\theta_{\rm obs},\phi_{\rm obs})
\end{align}
with ($\theta_{\rm obs}$,$\phi_{\rm obs}$) the angular coordinate of the observer measured in the orbit-centered frame described in Section \ref{misalignment}. Then we can see $\theta_{\rm obs} = i_{\rm orb}$, where $i_{\rm orb}$ is the orbital inclination to the line of sight, and $\phi_{\rm obs} = -\omega_p - \pi/2$. Hence, 
\begin{align}
\label{ylmo}
Y_{lm}(i_s,\phi_s) &= e^{- i m (\Omega_s t + \pi/2)} \sum_{m_o=-l}^{l} e^{-i m_o (\alpha_o - \pi/2)} d^l_{m_o m} (\beta_o)  Y_{l m_o} (i_{\rm orb}, -\omega_p-\pi/2) \nonumber \\
&= e^{- i m (\Omega_s t + \pi/2)} \sum_{m_o=-l}^{l} d^l_{m_o m} (\beta_o)  Y_{l m_o} (i_{\rm orb}, -\alpha_o - \omega_p)
&= 
\end{align}
Similar to the forcing terms, we can express this visibility term as an amplitude and phase shift, 
\beq
\label{ylmo2}
Y_{lm}(i_s,\phi_s) = e^{- i m (\Omega_s t + \pi/2)} V_{lm} e^{-i v_{lm}} \, ,
\eeq
with amplitude
\beq
\label{Vlm}
V_{lm} = \bigg[ \bigg( \sum_{m_o} d^l_{m_o m} Y_{l m_o}(i_{\rm orb},0) \cos \big[m_o (\alpha_o + \omega_p) \big] \bigg)^{\! 2} + \bigg( \sum_{m_o} d^l_{m_o m} Y_{l m_o}(i_{\rm orb},0) \sin \big[m_o (\alpha_o + \omega_p) \big] \bigg)^{\! 2} \bigg]^{1/2}
\eeq
and phase shift 
\beq
\label{v}
v_{lm} = {\rm atan2} \bigg[ \sum_{m_o} d^l_{m_o m} Y_{l m_o}(i_{\rm orb},0) \sin \big[m_o (\alpha_o + \omega_p) \big] , \sum_{m_o} d^l_{m_o m} Y_{l m_o}(i_{\rm orb},0) \cos \big[m_o (\alpha_o + \omega_p) \big] \bigg] \, .
\eeq

Inserting equation \ref{ylmo2} into equation \ref{dlumN} yields the observed luminosity variation due to a tidally forced mode,
\beq
\label{dlobs}
\frac{ \Delta L_{\alpha,N}}{L} = \frac{1}{2} \epsilon_l Q_{\alpha m} L_{\alpha m} V_{l m} X_{Nm} \frac{ \omega_{Nm} }{\omega_\alpha - \omega_{N m} + i \gamma_\alpha} e^{i(x_{Nm}-v_{lm})} e^{-i N \Omega t} \, .
\eeq
Note the time-dependence of the spin-term has canceled back out upon transformation to the observer's frame, so that tidally excited modes will be observed at integer harmonics of the orbital frequency regardless of the spin-orbit misalignment. 

Now, as above, we must sum over modes of different signs of $m$, $\omega_\alpha$, and $N$. This summation yields
\begin{align}
\label{dlobs2}
\frac{\Delta L_{|N|}}{L} \simeq \sum_{\omega_\alpha > 0, m \geq 0} & \epsilon_{l} X_{Nm} V_{lm} \big| Q_{\alpha m} L_{\alpha m} \big| \frac{\omega_{Nm}}{\sqrt{(\omega_{\alpha m} - \omega_{Nm})^2 + \gamma_{\alpha m}^2 }} \sin \big(N \Omega t + v_{lm}-x_{Nm} + \delta_{\alpha m} \big) \nonumber \\
&+ \epsilon_{l} X_{Nm} V_{lm} \big| Q_{\alpha -m} L_{\alpha -m} \big| \frac{\omega_{Nm}}{\sqrt{(\omega_{\alpha -m} + \omega_{Nm})^2 + \gamma_{\alpha -m}^2 }} \sin \big(N \Omega t + v_{lm}-x_{Nm} + \delta_{\alpha -m} \big) \nonumber \\
&+  \epsilon_{l} X_{N-m} V_{l-m} \big| Q_{\alpha -m} L_{\alpha -m} \big| \frac{\omega_{N-m}}{\sqrt{(\omega_{\alpha -m} - \omega_{N-m})^2 + \gamma_{\alpha -m}^2 }} \sin \big(N \Omega t + v_{l-m}-x_{N-m} + \delta_{\alpha -m} \big) \nonumber \\
&+ \epsilon_{l} X_{N-m} V_{l-m} \big| Q_{\alpha m} L_{\alpha m} \big| \frac{\omega_{N-m}}{\sqrt{(\omega_{\alpha m} + \omega_{N-m})^2 + \gamma_{\alpha m}^2 }} \sin \big(N \Omega t + v_{l-m}-x_{N-m} + \delta_{\alpha m} \big) \, .
\end{align}
In the aligned case, the first line represents prograde modes being forced in the same direction as the orbital motion, while the second line represents retrograde modes being forced in the direction of orbital motion. The third line represents retrograde modes being forced opposite to the direction of orbital motion, while the fourth line represents prograde modes being forced opposite to the direction of orbital motion. In the aligned case, the value of $X_{N-m}$ is very small and the third and fourth lines are negligible. However, these terms can be important for substantial misalignment. Note that $\omega_{Nm}$ can be positive or negative, but $\omega_{N-m} > 0$ and $\omega_\alpha > 0$ for all $N$ and $m$ in these sums. The first three terms in equation \ref{dlobs2} can produce resonantly excited modes, but the fourth term cannot (regardless of the spin-orbit alignment), so this term is less important. In the aligned limit, $V_{lm} \rightarrow Y_{lm}(i_s,0)$, $v_{lm} \rightarrow m(\alpha_o + \omega_p)$, $x_{Nm}-v_{lm} \rightarrow m( -\pi/2 - \omega_p) = m \phi_s$, and equation \ref{dlobs2} reduces to equation \ref{dlum3}.


\subsection{Determining Spin-Orbit Misalignment}

In practice, it can be difficult to observationally determine the spin-orbit misalignment angles $\alpha_o$ and $\beta_o$. However, they can be constrained if there is a measurement of the projected spin-orbit misalignment via the Rossiter-McLaughlin effect, or if there are measurements of rotation period, stellar radius, and $v \sin i_s$. We use a method similar to \cite{fabrycky:09} to present relations between $\alpha_o$, $\beta_o$, and observable quantities. 

In addition to the spin and orbit-oriented reference frames discussed above, we now consider the observer's reference frame with $\hat{z}_{\rm obs}$ pointing toward the observer. The $x_{\rm obs}-y_{\rm obs}$ plane is the plane of the sky, and we choose to orient this plane such that the $x_{\rm obs}-$axis lies on the intersection between the orbital plane and the plane of the sky. In this frame, the orbital axis has unit vector $ \hat{n}_{\rm orb} = \sin i_{\rm orb} \hat{y}_{\rm obs} + \cos i_{\rm orb} \hat{z}_{\rm obs}$. The stellar spin axis has unit vector $\hat{n}_s = \sin i_s \sin \lambda \hat{x}_{\rm obs} + \sin i_s \cos \lambda \hat{y}_{\rm obs} + \cos i_s \hat{z}_{\rm obs}$, where the angle $\lambda$ is the projected misalignment between the spin and orbital axis.

Let us also consider a reference frame defined relative to the orbit with $z_{\rm orb'}-$axis perpendicular to the orbital plane and $x_{\rm orb'}-$axis in the same direction as the $x_{\rm obs}$-axis above. In this frame, $\hat{n}_{\rm orb} = \hat{z}_{\rm orb'}$ and $\hat{n}_s = \sin \beta_o \sin (\alpha_o + \omega_p) \hat{x}_{\rm orb'} - \sin \beta_o \cos (\alpha_o + \omega_p) \hat{y}_{\rm orb'} + \cos \beta_o \hat{z}_{\rm orb'}$. The argument of periapse $\omega_p$ is the angle between $\hat{x}_{\rm orb}$ and $\hat{x}_{\rm orb'}$. The obs and orb' frames are related by a rotation about the $\hat{x}_{\rm obs} = \hat{x}_{\rm orb'}$ axis by an angle $i_{\rm orb}$, such that $\hat{y}_{\rm orb'} = \cos i_{\rm orb} \hat{y}_{\rm obs} - \sin i_{\rm orb} \hat{z}_{\rm obs}$, and $\hat{z}_{\rm orb'} = \sin i_{\rm orb} \hat{y}_{\rm obs} + \cos i_{\rm orb} \hat{z}_{\rm obs}$. Equating the unit vector of the spin axis in each frame thus requires
\beq
\sin i_s \sin \lambda = \sin \beta_o \sin (\alpha_o + \omega_p) \, ,
\eeq
\beq
\sin i_s \cos \lambda = -\sin \beta_o \cos (\alpha_o + \omega_p) \cos i_{\rm orb} + \cos \beta_o \sin i_{\rm orb} \, ,
\eeq
\beq
\label{cosis}
\cos i_s  = \sin \beta_o \cos (\alpha_o + \omega_p) \sin i_{\rm orb} + \cos \beta_o \cos i_{\rm orb} \, ,
\eeq
It is not generally possible to determine both $\alpha_o$ and $\beta_o$ from observed quantities because the above equations are not linearly independent. However, equation \ref{cosis} relates the values of $\alpha_o$ and $\beta_o$, provided that $i_s$, $i_{\rm orb}$, and $\omega_p$ can be measured,
\beq
\label{alpha}
\alpha_o = - \omega_p + \cos^{-1} \bigg[ \frac{ \cos i_s - \cos \beta_o \cos i_{\rm orb} }{\sin \beta \sin i_{\rm orb} } \bigg] \, .
\eeq

\section{Modes in the Traditional Approximation}
\label{trad}

Gravity modes in rapidly rotating stars are frequently studied in the traditional approximation (see \citealt{bildsten:96,lee:97,townsend:03} and references therein), which allows one to separate the radial and angular dependence of oscillation modes. Observed mode frequencies appear to conform very well to predictions of the traditional approximation \citep{moravveji:16,vanreeth:16}, although it fails to capture some dynamics of sub-inertial modes in convective zones \citep{mathis:14}.  The latitudinal depedendence of modes is no longer described by associated Legendre polynomials, but rather by Hough functions. The angular dependence of a mode $\alpha$ is found by solving Laplace's tidal equation:
\beq
\label{lap}
\mathcal{L}(H_{km}(\theta)) = - \lambda_k H_{km}(\theta),
\eeq
where $\lambda_k$ is an angular eigenvalue, $H_{km}$ is its associated Hough function, and the operator $\mathcal{L}$ is
\beq
\label{lop}
\mathcal{L} = \frac{\partial}{\partial \mu} \bigg( \frac{1-\mu^2}{1-q^2\mu^2} \frac{\partial}{\partial \mu} \bigg) - \frac{m^2}{(1-\mu^2)(1-q^2\mu^2)} + \frac{mq(1+q^2\mu^2)}{(1-q^2\mu^2)^2}
\eeq
with $\mu=\cos(\theta)$. The parameter $q$ that determines the behavior of the Hough functions is 
\beq
\label{q}
q = \frac{2 \Omega_s}{\omega}.
\eeq
Rotation becomes very important for $q \gtrsim 1$, and it is easily verified that the solutions to equation \ref{lap} converge to the associated Legendre polynomials as $q \rightarrow 0$. Similar to spherical harmonics, we normalize the Hough functions via 
\beq
\label{houghnorm}
\int dS H_{km}(\theta) H^*_{km}(\theta) = 1 \, .
\eeq
with the integral taken over a spherical surface.

Because the traditional approximation only works for g-modes, which exhibit small gravitational perturbations, it is almost always combined with the Cowling approximation, in which the gravitational perturbations of the modes are ignored. However, in tidal applications, the small (but finite) gravitational potential perturbations created by a g-mode determine the value of the overlap integral $Q_{\alpha}$ (equation \ref{Q}). While it is still possible to approximately calculate this overlap integral in the Cowling approximation using equation \ref{Q}, this equation is susceptible to numerical inaccuracies (see discussion in \citealt{fullerwd:11}), and we find it unsuitable to reliably calculate overlap integrals for low frequency g-modes.

Therefore, we choose to compute mode eigenfunctions using the traditional approximation but not using the Cowling approximation. To do this, we examine the linearized Poisson equation:
\beq
\label{poisson}
\frac{1}{r^2}\frac{\partial}{\partial r} \bigg( r^2 \frac{\partial}{\partial r} \delta \Phi(r,\theta,\phi) \bigg) + \nabla_\perp^2 \delta \Phi(r,\theta,\phi) = 4 \pi G \delta \rho(r,\theta,\phi),
\eeq
where $\delta \Phi(r,\theta,\phi)$ is the Eulerian gravitational potential perturbation, $\delta \rho(r,\theta,\phi)$ is the Eulerian density perturbation, and $\nabla_\perp^2$ is the angular part of the Laplacian. There arises an immediate problem with calculating gravitational potential perturbations of modes in the traditional approximation. Although $\delta \rho(r,\theta,\phi)$ has Hough function angular dependence, the quantity $\nabla_\perp^2 \delta \Phi(r,\theta,\phi)$ will in general not be described by a single Hough function.\footnote{The exception to this rule occurs when $q=0$, and the Hough function dependence of $\delta \rho$ reduces to an associated Legendre polynomial. Then, the angular operator $\mathcal{L} = r^2 \nabla_\perp^2$, and both $\nabla_\perp^2 \delta \Phi(r,\theta,\phi)$ and $\delta \rho(r,\theta,\phi)$ will be proportional to associated Legendre polynomials.} Consequently, the modes are not orthogonal because they will exhert gravitational torques on one another, and the radial and horizontal dependence of $\delta \Phi(r,\theta,\phi)$ are not separable.

In what follows, we ignore this issue and simply assume $\nabla_\perp^2 \delta \Phi(r,\theta,\phi)$ has the same Hough function angular dependence as $\delta \rho(r,\theta,\phi)$. We believe this approximation is justified for g-modes because $\delta \Phi$ is very small and does not strongly affect the mode behavior. For higher frequency modes (f modes and p modes), the approximation is also valid because they typically have $q \ll 1$ such that their angular dependence is nearly equal to a spherical harmonic. 

Letting $\delta \Phi(r,\theta,\phi) = \delta \Phi(r) H_{km}(\theta) e^{i m \phi}$, equation \ref{poisson} reduces to
\beq
\label{poisson2}
\frac{1}{r^2}\frac{\partial}{\partial r} \bigg( r^2 \frac{\partial}{\partial r} \delta \Phi(r) \bigg) + \frac{z_{km}}{r^2} \delta \Phi(r) = 4 \pi G \delta \rho(r),
\eeq
where
\beq
\label{zdef}
z_{km} = r^2\int dS H^*_{km}(\theta) \nabla_\perp^2 H_{km}(\theta)
\eeq
is an effective angular wave number. For $q \rightarrow 0$, $z_{km} \rightarrow -l(l+1)$. The Hough function dependence of $\delta \Phi$ changes the boundary conditions used to calculate the mode eigenfunctions. At both $r \rightarrow 0$ and $r \rightarrow R$, equation \ref{poisson2} reduces to (for non-radial modes)
\beq
\label{poisson3}
\frac{1}{r^2}\frac{\partial}{\partial r} \bigg( r^2 \frac{\partial}{\partial r} \delta \Phi(r) \bigg) + \frac{z_{km}}{r^2} \delta \Phi(r) \simeq 0.
\eeq
Letting $\delta \Phi \propto r^b$, we find $b = -1/2 \pm (1/2)\sqrt{1 - 4 z_{km}}$. Then regularity requires that
\beq
\label{inbc}
\frac{\partial}{\partial r} \delta \Phi = \frac{b_+}{r} \delta \Phi, \ \ {\rm at} \ r\rightarrow 0 
\eeq
and
\beq
\label{outbc}
\frac{\partial}{\partial r} \delta \Phi = \frac{b_-}{r} \delta \Phi, \ \ {\rm at} \ r\rightarrow R 
\eeq
where $b_+ = -1/2 + (1/2)\sqrt{1 - 4 z_{km}}$ and $b_- = -1/2 - (1/2)\sqrt{1 - 4 z_{km}}$. As $q \rightarrow 0$, we obtain the usual dependence $b_+ \rightarrow l$ and $b_- \rightarrow -(l+1)$. There are additional subtleties on the inner boundary condition for $\xi_r$ because the traditional approximation breaks down as $r \rightarrow 0$. In this study, we use the same approach described in \cite{fullerwd:14}.

With an explicit calculation of $\delta \Phi_\alpha$ for each mode, we may now use equation \ref{Q} to calculate the value of $Q_\alpha$ for each mode. Using the continuity equation and Poisson's equation, one can show
\begin{align}
\label{Q1}
\langle \xi_\beta | \bnab U \rangle &= h_{klm} R^2 \rho(R) \xi_r(R) U(R) + \int dV \delta \rho U \nonumber \\
&= \int dV \frac{1}{4 \pi G} U \nabla^2 \delta \Phi,
\end{align}
where we have assumed $\rho(R) \simeq 0$ to obtain the second line. Then, using Green's second identity, we have 
\begin{align}
\label{Q2}
\langle \xi_\beta | \bnab U \rangle &= \frac{1}{4 \pi G} \int dV \delta \Phi \nabla^2 U - \frac{R^2}{4 \pi G} \int dS \bigg( \delta \Phi \frac{\partial}{\partial r} U - U \frac{\partial}{\partial r} \delta \Phi \bigg)\bigg|_{r=R}
\end{align}
Then, since $\nabla^2 U =0$ for a tidal potential, we have
\begin{align}
\label{Q3}
\langle \xi_\beta | \bnab U \rangle &= \frac{R^2}{4 \pi G} \big(b_- - l\big) h_{klm} \delta \Phi(R) U(R),
\end{align}
where $h_{klm}$ describes the angular overlap between a Hough function and spherical harmonic,
\beq
\label{hdef}
h_{klm} = \int dS \, Y^*_{lm}(\theta,\phi) H_{km}(\theta) e^{i m \phi}.
\eeq
Then we find from equation \ref{Q} that 
\begin{align}
\label{Q4}
Q_{\alpha} = \frac{h_{klm} \big(b_- - l\big)}{4 \pi \omega_\alpha^2 } \delta \Phi_\alpha (R),
\end{align}
for all quantities expressed in dimensionless units ($G=M=R=1$), which reduces to $Q_\alpha = -(2l+1)\delta \Phi_\alpha(R)/(4\pi \omega_\alpha^2)$ in the non-rotating limit.

\section{Discussion and Conclusions}
\label{disc}

We have examined the visible luminosity oscillations produced by dynamical tides in eccentric binary systems known as heartbeat stars. The signature of dynamical tides is a stable oscillation at an exact integer multiple of the orbital frequency. Most observable tidally excited oscillations (TEOs) are produced by gravity modes (g modes) within one or both stars that are resonantly forced, i.e., a g mode frequency is nearly equal to an integer multiple of the orbital frequency. TEOs are expected to have higher amplitudes in hot stars with $T_{\rm ef} \gtrsim 6500 \, {\rm K}$, because the absence of surface convection zones in hot stars allows g modes to propagate much closer to the surface and produce larger surface temperature perturbations. In principle, it is straightforward to calculate the luminosity oscillations produced by TEOs given accurate stellar and orbital parameters. We have provided precise formulae to predict TEO amplitudes, frequencies, and phases, including Coriolis forces, non-adiabatic effects, and spin-orbit misalignment.

In practice, however, the uncertainties in stellar/orbital parameters make a precise calculation difficult because mode amplitudes are extremely sensitive to the resonant detuning between stellar oscillation frequencies and tidal forcing frequencies. It is often more constructive to compare observed and expected TEOs  using a statistical framework. In this approach, one can compare the number of observed TEOs exceeding a given threshold to the expected number of oscillations. One can also compare observed frequencies with the range in which TEOs are expected to be observed. This probabilistic approach assumes that tidal forcing frequencies (which occur at integer multiples of the orbital frequency) and oscillation mode frequencies are uncorrelated, i.e., the TEOs have no backreaction on the orbit. In reality, some feedback does occur because TEOs dissipate energy and cause tidal orbital evolution.

If a heartbeat star exhibits a very large amplitude TEO (e.g., KIC 8164262, see companion papers \citealt{hambleton:17,fullerkic81:17}) that is unlikely to stem from a chance resonance, it is a good candidate to be a resonantly locked mode. Resonantly locked modes arise from feedback between stellar evolution and TEOs, exciting a mode to large amplitude such that it increases the tidal dissipation rate, causing the orbital frequency to evolve such that the mode remains resonant. A detailed analysis of a population of heartbeat stars will determine whether resonance locking is a common phenomenon. If very few stars exhibit resonantly locked modes, this would indicate that resonance locks can be broken by some effect not accounted for here. Speculative possibilities include resonance disruption due to resonances of other modes, avoided crossings of g modes due to stellar evolution, non-linear mode coupling, or non-Keplerian orbital dynamics (e.g., precession or three-body effects).

If resonance locking does commonly occur in evolving systems, it can greatly enhance tidal dissipation rates. Moreover, it simplifies the physics of tidal dissipation, as orbital decay and spin synchronization proceed on stellar evolution time scales when resonance locking occurs. Resonance locking is not necessarily limited to eccentric binary star systems, and may operate in many astrophysical settings, including inspiraling WDs \citep{burkart:13}, planetary moon systems \citep{fullersattide:16}, and many other scenarios. We hope to investigate new possibilities in future works.

\section*{Acknowledgments}

I am grateful to Susan Mullally for providing the data in Figure 1, and to the anonymous referee for a thoughtful report. I thank Rich Townsend and Zhao Guo for helpful discussions. This research was supported in part by a Lee DuBridge Fellowship at Caltech,  the National Science Foundation under grants AST-1205732 and PHY-1125915, and the Gordon and Betty Moore Foundation through grant GBMF5076.

\bibliography{../Work/references/fuller,../Work/references/heartbeat,../Work/references/neutronstars,../Work/references/massivestars,../Work/references/angmomtrans,../Work/references/astero}

\appendix

\section{MESA Models}
\label{mesa}

Our stellar models are made using the MESA stellar evolution code \citep{paxton:11,paxton:13,paxton:15}, version 9575. Important settings include the use of  convective overshoot with an exponential decline above the convective zone, and a small amount of molecular diffusivity of $D = 1 \, {\rm cm}^2{\rm s}^{-1}$. Our effective overshooting parameter of $f = 0.01$ is similar to values inferred for slowly pulsating B stars \citep{moravveji:15,moravveji:16} and solar-like oscillators \citep{deheuvels:16}. Larger diffusivities smooth the compositional gradient and its contribution to the Brunt-V\"{a}is\"{a}l\"{a} frequency, causing less mode trapping near the core and smoothing out the dips in Figures \ref{ModeL}. An inlist for our models is given below. 

\begin{verbatim}

&star_job  

      pgstar_flag = .true.
      
/ ! end of star_job namelist

&controls

      write_pulse_data_with_profile = .true.
      pulse_data_format = 'GYRE'
      
      initial_mass = 1.7
      initial_z = 0.02
      use_Type2_opacities = .true.
      Zbase = 2.d-2 

      cool_wind_RGB_scheme = 'Reimers'
      cool_wind_AGB_scheme = 'Blocker'
      RGB_to_AGB_wind_switch = 1d-4
      Reimers_scaling_factor = 0.5
      Blocker_scaling_factor = 0.5
     
      overshoot_f_above_nonburn_core = 0.01
      overshoot_f0_above_nonburn_core = 0.005
      overshoot_f_above_nonburn_shell = 0.01
      overshoot_f0_above_nonburn_shell = 0.005
      overshoot_f_below_nonburn_shell = 0.01
      overshoot_f0_below_nonburn_shell = 0.005

      overshoot_f_above_burn_h_core = 0.01
      overshoot_f0_above_burn_h_core = 0.005
      overshoot_f_above_burn_h_shell = 0.01
      overshoot_f0_above_burn_h_shell = 0.005
      overshoot_f_below_burn_h_shell = 0.01
      overshoot_f0_below_burn_h_shell = 0.005

      set_min_D_mix = .true.
      min_D_mix = 1d0
     
      photo_interval = 100
      profile_interval = 3
      max_num_profile_models = 3000
      history_interval = 3
      terminal_interval = 1
      write_header_frequency = 10
      max_number_backups = 500
      max_number_retries = 1000
      max_timestep = 1.15d14  ! in seconds.    

      mesh_delta_coeff = 0.3
      varcontrol_target = 5.d-4

      xa_central_lower_limit_species(1) = 'h1'
      xa_central_lower_limit(1) = 0.001 
      
/ ! end of controls namelist

\end{verbatim}

After constructing stellar models, we compute non-adiabatic oscillation mode properties with the GYRE pulsation code \citep{townsend:13}, version 5.0. We employ rotation via the traditional approximation. An inlist is given below.

\begin{verbatim}

&model
   model_type = 'EVOL'  ! Obtain stellar structure from an evolutionary model
   file = 'profile118.data.GYRE'    ! File name of the evolutionary model
   file_format = 'MESA' ! File format of the evolutionary model
   uniform_rot = .true. !Turn on rotation
   Omega_units = 'RAD_PER_SEC' !Turn on rotation
   Omega_rot = 2.433e-5
/

&mode
   l = 2                     ! Harmonic degree
   m = 2
/

&osc
   outer_bound = 'ZERO' ! Use a zero-pressure outer mechanical boundary condition
   rotation_method = 'TAR' !Use traditional approximation
   nonadiabatic = .TRUE.
/

&num
   diff_scheme = 'COLLOC_GL2' ! 2nd-order Magnus solver for initial-value integrations
   n_iter_max = 100
/

&scan
   grid_type = 'INVERSE' ! Scan for modes using a uniform-in-period grid; best for g modes
   freq_min = 0.1     ! Minimum frequency to scan from
   freq_max = 10.0        ! Maximum frequency to scan to
   n_freq = 1000          ! Number of frequency points in scan
   grid_frame = 'COROT_O'
   freq_frame = 'COROT_O'
/

&grid
   alpha_osc = 10
   alpha_exp = 2
   n_inner = 10
/

&ad_output
   summary_file = 'summary_ad.txt'   ! File name for summary file
   summary_file_format = 'TXT'   ! Format of summary file
   summary_item_list = 'M_star,R_star,l,n_pg,omega,E_norm,f_T,psi_T'   ! Items to appear in summary file                     
   mode_file_format = 'TXT'  ! Format of mode files
   mode_item_list = 'l,n_pg,omega,x,xi_r,xi_h,eul_phi,lag_L,lag_T,rho,Omega_rot'  ! Eigenfunction outputs
/

&nad_output
   summary_file = 'summary_nad.txt'  ! File name for summary file
   summary_file_format = 'TXT' ! Format of summary file
   summary_item_list = 'M_star,R_star,l,n_pg,omega,E_norm,f_T,psi_T'  ! Items to appear in summary file                      
   mode_template = 'mode%n.txt' ! File-name prefix for mode files
   mode_file_format = 'TXT'   ! Format of mode files
   mode_item_list = 'l,n_pg,omega,x,xi_r,xi_h,eul_phi,lag_L,lag_T,rho,Omega_rot'  ! Eigenfunction outputs
/

\end{verbatim}

We implement a couple important changes when computing our stellar oscillation modes. First, we renormalize GYRE's eigenfunctions according to equation \ref{modeorthrot}. Second, we set $c_{\rm rad}' = -3$ using the statement \verb|dc_rad = -3._WP| in GYRE's oscillation equations. The purpose of this is to eliminate terms arising from gradients in the fraction of energy carried by radiation, $c_{\rm rad}$. We find these terms become very large in the partial hydrogen convective zone near the surfaces of stars with $T_{\rm eff} \sim 7000 \, {\rm K}$, due to the small scale height and sharp dependence of opacity on temperature. Consequently, the radial derivative of the luminosity perturbation eigenfunction, $\delta L/L$, becomes very large in this region. We deem this behavior unphysical, because the thermal time in these regions is much shorter than g mode pulsation frequencies. In this regime, the oscillations should be roughly isothermal (rather than adiabatic), and the luminosity perturbation should be essentially frozen in these layers, i.e., the derivative of $\delta L/L$ should be near zero. 

We find that implementing the change above yields more physical eigenfunctions. The ultimate source of the problem is that GYRE does not compute changes to the convective flux perturbation, which is a well known and very difficult problem in asteroseismology \citep{unno:89}. Our implementation effectively sets the convective luminosity perturbation equal to zero, while GYRE's current default implementation effectively sets the convective entropy perturbation equal to zero (see equations 21.6 and 21.7 of \citealt{unno:89}). Neither method is physically accurate. However, after experimenting with different approximations for convective flux/entropy perturbations, we find that many prescriptions (including the one we adopt) yield somewhat similar results, with slight differences in the surface luminosity perturbation in Figure \ref{ModeL}. Based on these experiments, we believe our computed values of $L_\alpha$ are accurate at the factor of 2 level for modes with $f_{\alpha} \sim 0.5 \, {\rm c/d}$ in stars with $T_{\rm eff} \sim 7000 \, {\rm K}$. For higher frequency modes and stars of cooler/warmer temperatures, convective flux perturbations are less of an issue, and our values of $L_\alpha$ are probably significantly more reliable. GYRE's default prescription produces very different (and in our opinion, unphysical) eigenfunctions due to the large $c_{\rm rad}'$ term as described above. Since g modes have very little inertia in the hydrogen partial ionization region, GYRE's oscillation mode frequencies and growth rates are likely to be insensitive to these effects. However, the surface luminosity perturbation is significantly impacted, and future work should attempt to more accurately treat convective flux perturbations in these regions.

\end{document}